\documentclass[twocolumn,showpacs,preprintnumbers]{revtex4-1}

\usepackage{graphics} 
\usepackage{graphicx} 
\usepackage{amssymb} 
\usepackage{amsthm} 
\usepackage[english]{babel} 
\usepackage{amsmath} 
\usepackage{siunitx} 
\usepackage{dcolumn} 
\usepackage{longtable} 

\usepackage{changepage} 

\begin{document}


\title{Extrapolations of nuclear binding energies from new linear mass relations}
\author{D. Hove, A.S. Jensen, K. Riisager}
\affiliation{Department of Physics and Astronomy, Aarhus University, DK-8000 Aarhus C, Denmark} 
\date{\today}


\begin{abstract}

We present a method to extrapolate nuclear binding energies from known
values for neighbouring nuclei. We select four specific mass relations
constructed to eliminate smooth variation of the binding energy as
function nucleon numbers. The fast odd-even variations are avoided by
comparing nuclei with same parity.  The mass
relations are first tested and shown to either be rather accurately
obeyed or revealing signatures of quickly varying structures.
Extrapolations are initially made for a nucleus by applying each of
these relations.  Very reliable estimates are then produced either by
an average or by choosing the extrapolation where the smoothest
structures enter. Corresponding mass relations for $Q_{\alpha}$ values
are used to study the general structure of super-heavy elements. A minor neutron shell at $N = 152$
is seen, but no sign of other shell structures are apparent in the
super-heavy region. Accuracies are typically substantially better than
$0.5$~MeV. 
\end{abstract}


\maketitle



\section{Introduction \label{Introduction}}

The importance of accurate knowledge of nuclear masses is not disputed
by anybody.  Unprecedented numbers of precise measurements are
available \cite{lun03,blo12}, but many particle stable masses are still
unknown. The masses are collected in comprehensive mass tables
\cite{aud03} which also contain estimates based on smooth
extra/interpolations and consistency between a number of related
particle and cluster separation energies.

Different types of theoretical models are also used to estimate and
predict nuclear masses of interest. They are almost all at some point
employing phenomenological parametrization. The original example is
the semi-classical mass formula by von Weiz\"{a}cker and Bethe
\cite{boh98}, where four parameters are fitted to known masses and all
others can be predicted. Much more sophisticated versions are
developed where the same idea of expanding in terms of neutron and
proton numbers systematically is exploited in the liquid droplet model
\cite{Myers,mol12}.

The success of the liquid drop models is due to the overall continuous
behaviour of nuclear masses as function of neutron and proton numbers,
and of course on the inclusion of the correct physics ingredients of
volume, surface, Coulomb and symmetry terms. After the bulk part of
the nuclear masses are described the smaller contributions are
highlighted as the remaining part. This is much more difficult to
describe as the origin is in a number of very different correlations
expressed as e.g. shell effects, deformations, and pairing. These three
correlations occur rather systematically and can to some degree be
accounted for in the droplet models. However, the severe limitation is
that predictions beyond the experimentally known regions quickly
become rather inaccurate.  

Improvement in predictive power is obtained by microscopic mean-field
models, i.e. Hartee-Fock-Boguliubov, Density Functional Theory and
Thomas-Fermi calculations \cite{gor08,erl12}. Now the phenomenology
enters as the nucleon-nucleon interactions used as input, and
determined from general symmetry principles and by fitting to
resulting computed properties. Here the self-consistency is necessary
to have reliability beyond the fitted regions. At some level the
liquid drop bulk properties must be reproduced if these models are to
be successful. This is more directly exploited in the micro-macro
models where the microscopic fluctuating part first is extracted from
a mean-field shell model computation and the average smooth part is
replaced by liquid drop expressions \cite{bra97}.

The origin of nuclear masses is the nucleon-nucleon interaction which
implies that the different nuclei have (perhaps complicated) related
masses. This is explored in ab initio calculations of nuclear masses
from the basic interaction \cite{nav09}. It is exploited in a
completely different way in a series of so-called mass relations where
Garvey-Kelson is the most well-known \cite{Garvey}. It is based on
counting the number of pairwise interactions in different nuclei and by
adding for example three mass difference between two nuclei, the result
should be zero. This is tested to be true for known masses with an
average accuracy of about 500~keV \cite{men08}.

It is then interesting to test whether the previous mass formulae obey
the rather accurate Garvey-Kelson mass relations. This turns out to be
essentially true for the measured masses, but as soon as
extrapolations are involved the accuracy drops by about a factor of
two \cite{bar08}. The phenomenology is only really trustworthy within
the fitted region. There is apparently one exception in the
Duflo-Zuker mass formulae constructed from the same principles as the
Garvey-Kelson mass relations \cite{bar08,duf95}.

A different principle was used in extraction of pairing properties
where emphasis rather than cancellation is desired. Odd-even mass
differences between neighbouring nuclei already reveal these effects.
An improvement is obtained in the slightly more complicated
combination where an average of the two neighbouring mass differences
is used \cite{boh98}. A further extension to include more masses led
Jensen et al. \cite{Aksel} to formulate mass relations obeying a
general principle of cancellation of all smooth terms up to any
desired order. The practical choice is second order, since the
necessary nuclei otherwise may differ too much. It is interesting to
note that Garvey-Kelson relations also eliminate all smooth terms up
to second order. Although never emphasized previously, this is
obviously a convincing reason for their success.

Different mass combinations can now be chosen to either emphasize
specific correlations or to avoid them for example by cancellation.
The latter choice produces a combination of masses equal to zero, which
means any of these masses can be expressed as a linear combination of
the other ones. Thus, if correct such mass relations are directly
applicable for one-step extrapolations beyond known mass regions.
Similar extrapolations can be made with Garvey-Kelson mass relations
but they do not allow special choices where for example odd-even
effects a priori are absent or emphasized. Other correlations could be investigated as
well if a mass combination can be found to highlight them. 

It has been suggested that nuclear masses have a component of chaotic
behaviour amounting to $2.78/A^{1/3}$~MeV \cite{mol06} which amounts to
between $0.5$ and $1.4$~MeV. This seems to be an exaggeration as suggested by the observation that specific regions exhibit (unknown)
correlations \cite{olu06} accounting for maybe half of this
amount. This is also indicated by the rather small root mean square deviation of less than $100$~keV obtained by the $12$ point Garvey-Kelson mass interpolation \cite{bar08}. Thus, any mass extrapolation can ultimately only meaningfully aim for an
accuracy of at most $200-300$~keV with global mass formulae.

The purpose of this paper is to present four linear mass relations
between isotopes capable of removing smoothly varying
contributions. When applied to isotopes with measured binding
energies, these mass combinations should have a tendency to cancel
completely barring influences from other significant contributing
factors. Expressing unknown masses as linear combinations of known ones should allow for the extrapolating of these unknown masses. This will all be based on isotopes in their ground state
configurations. The relatively few assumptions needed to establish the
fundamental model is the greatest advantage of the method.  As a result all conclusions will be based purely on combinations
of binding energies, without the need for other theoretical
considerations.

Our focus will be divided between extrapolating unknown binding
energies and studying the structure of the super-heavy elements.  We
shall use the method introduced in \cite{Aksel} to construct mass
relations.  Here it will not be attempted to verify the existence and
scale of the effects that influence the binding energy. We shall use
suitable mass relations to eliminate all or most of the systematic
dependencies of the binding energy on nucleon numbers. The legitimacy
of the elimination will be apparent from the results of applying the
mass relations.

The fundamental model, along with the argumentation supporting it,
will be presented in sect.\ (\ref{Sect. mass rel}). The majority of
the necessary formulations will be included there as well. It is then
possible to define the specific mass relations needed for the
applications, and this is also included in sect.\ (\ref{Sect. mass
  rel}).

Applying these mass relations individually with the purpose of
extrapolating to new binding energies is done in
sect.\ (\ref{Sect. Indi1}) and (\ref{Sect. Indi2}). In
sect.\ (\ref{Sec. Q}) the mass relations are used with $Q_{\alpha}$
values. This has a number of advantages. In particular, it is possible
to examine the region of super-heavy isotopes in greater detail. The
$Q_{\alpha}$ values are very useful in analysis of general structures
appearing in the binding energy. By comparing extrapolations from the
individual mass relations it is possible to calculate more precise
results either by simple averages or by choosing the smoothest
extrapolation. Such combinations are presented in
sect.\ (\ref{Sec. Com. Res.}) along with the numerical results in
table \ref{Tab. Ex}. Finally, sect.\ (\ref{Conclusion}) contains a
brief summary and the conclusions.



\section{The mass relations \label{Sect. mass rel}}

The idea behind the mass relations is that the nuclear many-body systems
all are derived from the same basic interactions, and hence different
nuclei should have related binding energies. Various principles are
applied for different mass relations. We shall focus on one type
where we first describe the general principles, then we derive some
useful properties, and finally we specify the applications in the last
subsection.

\subsection{General assumptions \label{Sect. B engy}} 

The mass formula is often divided into a sum of three different types
of terms. First the dominating term, $B_{LD}(N, Z)$, describing the
smoothly varying gross properties of the binding energy as function of
neutron and proton numbers, $N$ and $Z$. This is the liquid drop, or
droplet, model with the classical four terms, that is volume-,
surface-, Coulomb- and symmetry-energy. The specific form and the
precise numerical values are not important since the smooth
character is only necessary to eliminate unwanted contributions. This is
achieved through suitable linear combinations of the nuclear binding
energies as elaborated in sect.\ (\ref{Sect. Manipulating}).

Second, a term accounting for shell effects, $B_{sh}(N, Z)$, arising
from quantum mechanical correlations favouring special (spherical)
configurations. Third, a term, $\Delta (N, Z)$, describing systematic
but not smoothly varying contributions to the binding energy. This
can be odd-even and other similar (short-ranged) correlation effects.  In
total, we have the binding energy separated into such distinct terms,
where each is a function of the nucleon numbers $N$, and $Z$:
\begin{align}
B(N, Z) = B_{LD}(N, Z) + B_{sh}(N, Z) + \Delta (N, Z) \; .\label{B}
\end{align}
Explicit addition of terms describing other effects, for instance the
possible tendency to form $\alpha$-particles within nuclei, could also
be included. However, the possible nature of $\alpha$-clusters is
presently not our prime focus, and furthermore the energy gain from
these clusters are also very small or possibly very smoothly varying
\cite{Aksel}. 

Since the existence of both neutron and proton shells is undeniable,
the second term, $B_{sh}(N, Z)$, is an inescapable necessity. The
major shells are prominent only in relatively narrow regions of
nucleon numbers. A slowly varying contribution from
$B_{sh}$ between shells can then essentially be eliminated by
the same procedure as $B_{LD}(N, Z)$. This claim will be
substantiated by the results in sect.\ (\ref{Sect. Ext}).

The systematic third term, $\Delta (N, Z)$, is more complicated since
it is composed of several effects. It includes three different pairing
effects, along with the Wigner term related to the isospin symmetry,
all of which are more subtle in nature than the smooth term. However,
they are all smooth functions of nucleon numbers provided isotopes with same parity are compared and the $N=Z$
line is not crossed. We shall in this paper impose these restrictions
on the employed extrapolations, although we expect to encounter
occasional signals of these terms.

The terms in eq.\ (\ref{B}) do not necessarily constitute a complete
expression for the binding energy. Additional overlooked
or unknown effects might also contribute in different ways. However, we
expect that any such neglected but significant effects will produce a
clear deviation from the systematic results, and thereby reveal
itself. This will be considered in relation to the actual numerical
results presented in sect.\ (\ref{Sec. Com. Res.}).

\subsection{Manipulating the binding energy \label{Sect. Manipulating}}

A flexible method to manipulate binding energies was discussed in
\cite{Aksel} with the aim of isolating specific contributing effects.
A possibility is then to study individual effects in relative
isolation. However, this flexibility also indirectly enables the
extrapolation of unknown binding energies. The idea is to combine
separation energies in a manner reminiscent of a second order
difference.
\begin{align}
Q(n_1, z_1; n_2, z_2) =& -S(N - n_1, Z - z_1) + 2 S(N, Z) \notag \\ 
&- S(N + n_1, Z + z_1) \;.  \label{Q}
\end{align}
The separation energy of $n_2$ neutrons and $z_2$ protons in any isotope is given as a difference between binding energies.
\begin{align}
S(N, Z) = B(N, Z) - B(N - n_2, Z - z_2) \;.  \label{S}
\end{align} 
Calculating $Q$ using eqs.\ (\ref{B}) and (\ref{S}) results in an
expression for $Q$, which like the original expression for $B$ in
eq.\ (\ref{B}), can be separated in three terms, i.e.
\begin{align}
Q = Q_{LD} + Q_{sh} + Q_\Delta \;.  \label{QS}
\end{align}

Depending on the chosen $(n_1, z_1; n_2, z_2)$, some terms will be
diminished while others will be emphasized. The contributions from
the last two terms in eq.\ (\ref{QS}) vary greatly in size depending
on the chosen $(n_1, z_1; n_2, z_2)$, but common for all
configurations is the fact that the smooth terms are almost completely
eliminated. Interpreting the discrete variables $N$ and $Z$ as
global, continuous variables automatically results in an elimination
up to and including the second order in the Taylor expansion of the
smooth terms around $(N, Z)$. The leading order contribution to
continuous functions, $\tilde{B}$ and $\tilde{Q}$ analogous to $B$ and
$Q$, is then third order in the Taylor expansion, that is
\begin{align}
\tilde{Q} =& 
- \frac{\partial^3 \tilde{B}}{\partial N^2 \partial Z} n_1 (n_1 z_2 + 2 n_2 z_1)
- \frac{\partial^3 \tilde{B}}{\partial N^3} n_1^2 n_2 \notag \\ 
&- \frac{\partial^3 \tilde{B}}{\partial Z^3} z_1^2 z_2 
- \frac{\partial^3 \tilde{B}}{\partial N \partial Z^2} z_1 (n_2 z_1 + 2 n_1 z_2)
 \; , \label{3orde}
\end{align}
as seen by direct expansion. This remaining contribution will always
be present for smooth functions when mass relations based on
eq.\ (\ref{Q}) are constructed. It either has to be corrected for or
included in accuracy estimates.

By severe reduction of the smooth contributions to a size like in
eq.\ (\ref{3orde}) other effects would stand out. Desired effects
can then be emphasized by suitably chosen configurations
$(n_1,z_1;n_2,z_2)$. The shell effect in particular will figure
prominently in certain parts of the nuclear chart, and the validity of
some extrapolations in these areas will therefore be more
doubtful. However, lacking an accurate expression for the general
contributions from shell effects, it is difficult to construct
appropriately corrected mass relations. Also any expression
describing shell effects would be another source of error in the
extrapolations. Thus, we shall not attempt to account for the shell
effects, although perhaps detect their presence by observing
systematic deviations.

Instead of the separation energies in eq.\ (\ref{Q}) we can use
similar combinations arising from $Q_{\alpha}$ values, that is for
$n_2=z_2=2$
\begin{align}
S(N, Z) - B(^4\mathrm{He}) 
= -Q_{\alpha} \;.
\end{align}
The advantage is that $Q_{\alpha}$ sometimes is much more accurately
known than nuclear masses themselves, and this is especially
pronounced for super-heavy nuclei. This observation is very well
established by the experimental techniques where masses are measured
relative to other masses. Then it is possible to use eq.\ (\ref{Q})
with $Q_{\alpha}$ values, which leads to
\begin{align}
Q
=& \, Q + B(^4\mathrm{He}) - 2 B(^4\mathrm{He}) + B(^4\mathrm{He}) \notag \\
=& \, Q_{\alpha}(N - n_1, Z - z_1) - 2 Q_{\alpha}(N, Z) \notag \\ 
&+ Q_{\alpha}(N + n_1, Z + z_1) \;. \label{Qalph}
\end{align}
Other types of conclusions may then become possible from $Q_{\alpha}$
relations, as, in addition to the better accuracy, only three
measured values enters eq.\ (\ref{Qalph}) in contrast to the
four terms arising from eqs.\ (\ref{Q}) and (\ref{S}).

\subsection{Constructing specific mass relations \label{Sec. Cal mass}}

The aim is to find a reliable extrapolation of binding energies
through the mass relations in sect.\ (\ref{Sect. Manipulating}). This
is accomplished by carefully choosing the configuration in
eq.\ (\ref{Q}) such that the result ideally is zero. If in a certain
area of the nuclear chart, limited only by the available measured
isotopes, a mass relation is prone to return the value zero, it is
reasonable to assume this tendency would continue beyond the known
isotopes. Unknown binding energies can then be calculated directly
from a given mass relation. However, such extrapolation is only
reliable if the chosen mass relation in fact eliminates all
contributions from the binding energy in eq.\ (\ref{B}). Even then
care has to be taken to avoid outlandish results.

Many mass formulas
have a tendency to deviate significantly when extrapolating outside
the experimentally known region \cite{bar08}. The present method does not rely on
a specific form of a mass formula. However, eq.\ (\ref{Q}) allows for
an endless number of possible mass relations by choosing $(n_1, z_1;
n_2, z_2)$ accordingly, and using too great values for $n_i$ and $z_i$
would make the approximation of eq.\ (\ref{Q}) as a derivative less
accurate. The likelihood of combining different effects in the result
increases when combining isotopes farther apart, and the extrapolation
would also be less accurate. Similarly, the mass relation could be
chosen to eliminate the smooth parts to any order desired.
Unfortunately, this would also come at the expense of reliability
since isotopes farther apart would be required.

\begin{figure}
\centering
	\includegraphics[width=\columnwidth]{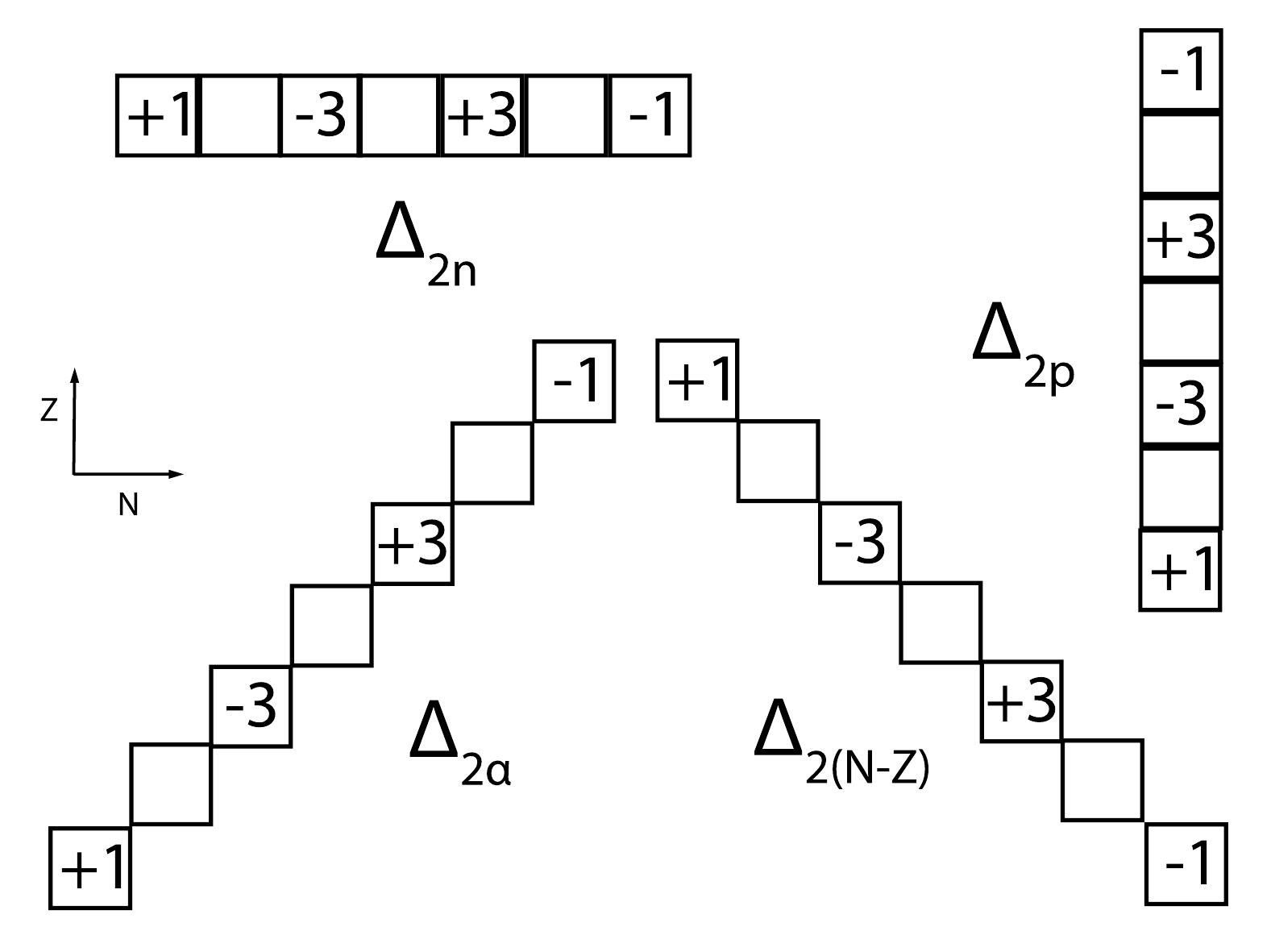}
\caption{The physical structures of the four mass relations in eq.\ (\ref{Constructions}). The weights assigned to the different isotopes reflect the concatenation of the six terms from eqs.\ (\ref{Q}) and (\ref{S}) into four. \label{Structures.pdf}}
\end{figure}

We therefore only apply mass relations where $n_i$ and $z_i$ never are
larger than 2. Furthermore, to avoid the quickly varying pairing
contribution, we choose to compare nuclei of the same odd-even
character. In total we use here four mass relations where $n_i$ and
$z_i$ are $0$ and $2$. They combine nuclei with fixed $N$, $Z$,
$A=N+Z$, and $N-Z$, respectively, that is defined by
\begin{align}
\Delta_{2n}(N,Z)      &= Q(2, 0; 2, 0)  \notag \\ 
\Delta_{2p}(N,Z)      &= Q(0, 2; 0, 2)  \notag \\ 
\Delta_{2\alpha}(N,Z) &= Q(2, 2; 2, 2)  \notag \\ 
\Delta_{2(N-Z)}(N,Z)  &= Q(2, -2; 2, -2) \; . \label{Constructions}
\end{align}
The actual nuclei in these four mass relations can be seen in
fig.~\ref{Structures.pdf} where the original six nuclei from
eq.\ (\ref{Q}) and (\ref{S}) reduce to only four with different
weights. These four mass relations should, ideally, completely
eliminate any contributions from pairing effects. Of course, the
actual results will not be so idealized, and will at the very least
include remnants of the smooth term. Some minor pairing contribution
might still remain, since no systematic theory can account for all
these effects as discussed by Friedman \textit{et
  al}.\ \cite{Friedman}. Still, the combinations shown in
fig.~\ref{Structures.pdf} seem intuitively to be more likely to add
up to zero, and thereby providing useful mass relations for the
extrapolations.

\section{Extrapolations \label{Sect. Ext}}

The actual analysis is divided into two subsections. First the results
are examined individually from applying the four mass relations to the
available measurements of binding energies. The tendencies are
discussed to emphasize the relevant structures and provide insight
into the viability of the general method. The areas accessible to the
mass relations will also be determined in the process. The measurements are from Audi and Meng \cite{Audi} for isotopes in their ground state with nucleon number, $A$, spanning $0-295$.

Second, the mass relations are applied to $Q_{\alpha}$ values with the purpose of analysing general structures found among the super-heavy elements. The measurements of $Q_{\alpha}$ are also from Audi and Meng \cite{Audi}, unfortunately, they are not necessarily of isotopes in their ground state.

\subsection{Procedure and general behavior \label{Sect. Indi1}}

The general method described in sect.\ (\ref{Sec. Cal mass}) is
idealized, and constitutes the simplest and most obvious way to
perform the extrapolations. However, a slightly more complicated
procedure is applied to increase accuracy and estimate uncertainty. The
fundamental idea is still to combine four different isotopes either
horizontally, vertically, or diagonally.

First the mass relation is tested locally, that is with $\Delta_{2n}$
as the example and $(N+2,Z)$ as the unknown we compute $\Delta_{2n}(N-2,Z)$, $\Delta_{2n}(N-4,Z)$,
and $\Delta_{2n}(N-6,Z)$. Each would be zero if the mass relation is
exactly obeyed.  A systematic tendency in the region can be accounted
for by computing the non-zero average value which is used for
$\Delta_{2n}(N,Z)$ in the prediction of the unknown $(N+2,Z)$ binding
energy. Obviously, a systematic tendency is then accounted for in the
prediction which furthermore now has an extrapolation uncertainty
attached from the spread around the average value of the mass
relation.

With the diagonal relations ($\Delta_{2(N-Z)}$ and $\Delta_{2\alpha}$)
it is impossible to calculate an average based on three preceding
values. This would reduce the available extrapolations almost to
none. Instead only the immediately preceding value is used for
$\Delta_{2(N-Z)}$.

The uncertainties of the actual extrapolations have two general
sources. The uncertainties in the measurements combine with the
uncertainty of the predicted (non-zero) average value.  Since this
expected average is based on three different, but overlapping
applications of the mass relation, this statistical error is the
combination of six different uncertainties in measurements. Depending
on the specific isotopes, and how well they have been measured, this
uncertainty can at times be very significant.

Recently, it has been suggested by Olofsson \textit{et
  al}.\ \cite{olu06} that the distribution of binding energies
inherently is, at least partly, chaotic in nature. This is still
subject to discussion as Molinari and Weidenm\"{u}ller \cite{mol06}
interpret the results as being due to residual interactions in the
shell model. However, to account for any (chaotic) fluctuations the
variation of the average value, computed from the three mass
combinations, must be included in the final uncertainty of the
extrapolated value.

To achieve this we combine the two different contributions to the
uncertainty, that is from measurement and average. Thus, $r_i \pm
\sqrt{s_i^2 + v_i^2} = r_i \pm \sigma_i$, where $i$ labels the applied
relation, $r_i$ is the extrapolated value, $s_i$ is the measurement
uncertainty, $v_i$ is the variation, and $\sigma_i$ is the final
uncertainty of the extrapolation. The applicability of this
extrapolation method has limits, and some energies cannot be
meaningfully extrapolated. Consequently, only results where
$\sigma < 500 \, \si{\kilo\electronvolt}$ are included, since otherwise the
extrapolated values are too uncertain to be of interest.

\begin{figure*}
\centering
	\includegraphics{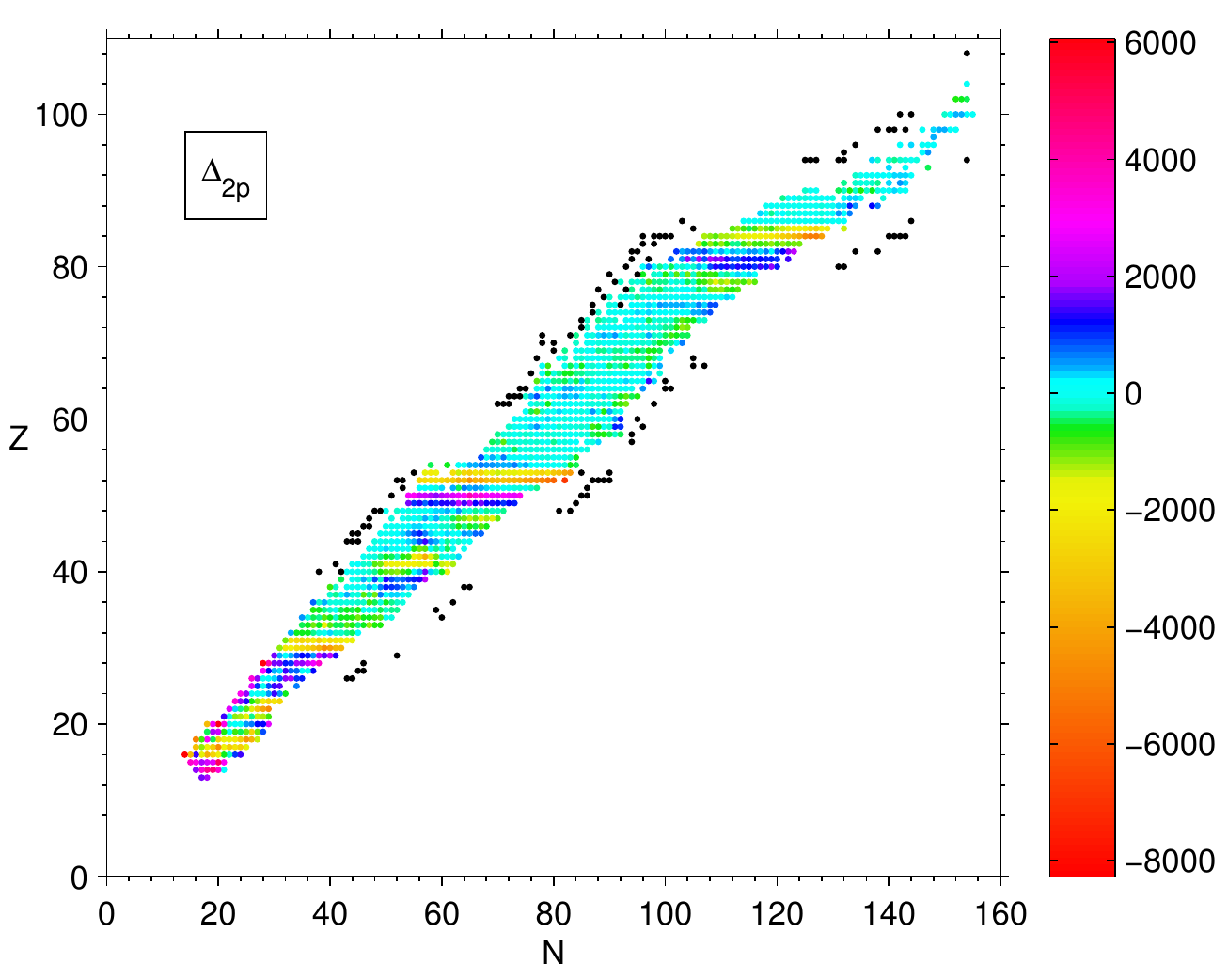}
	\includegraphics{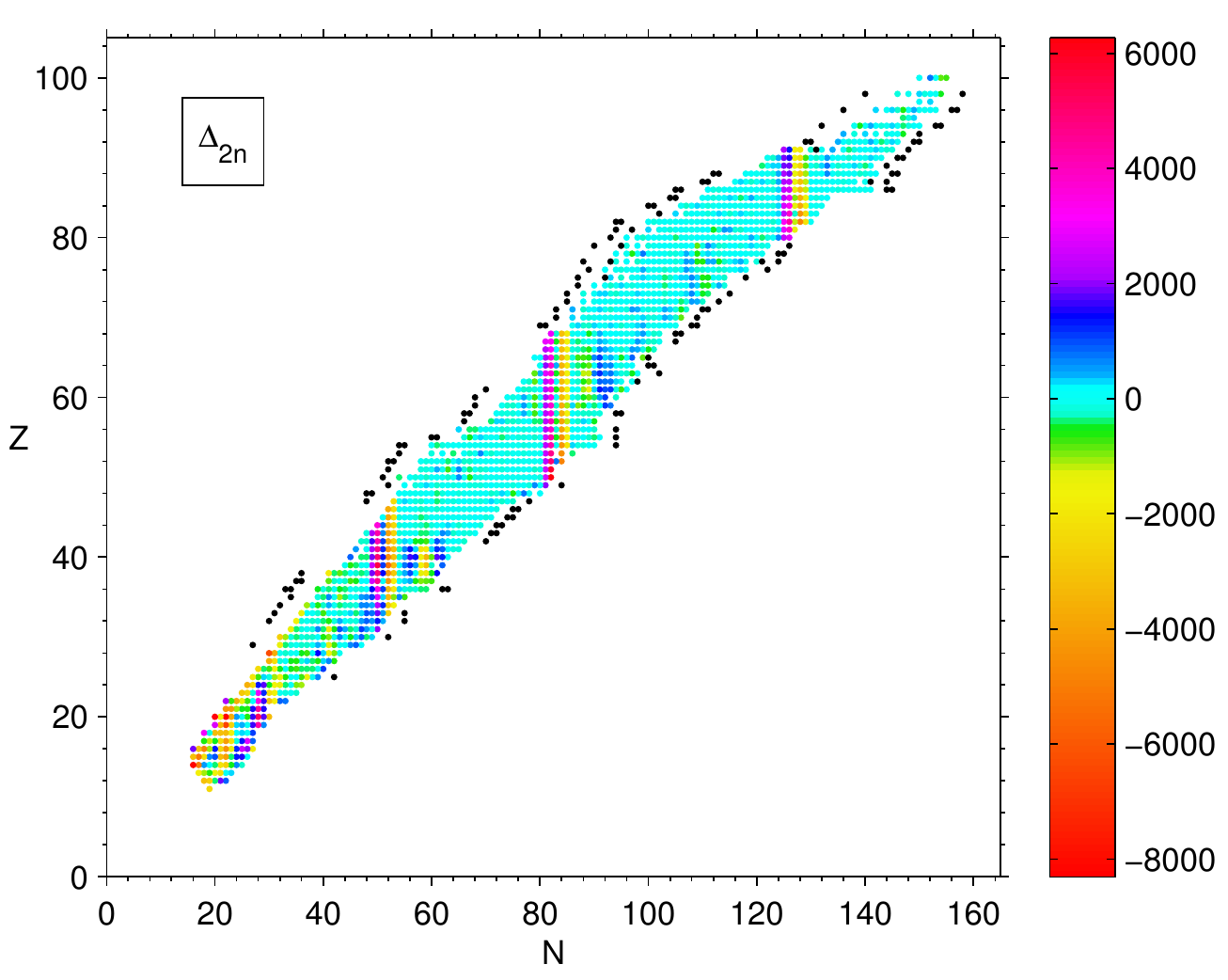}
\caption{The vertical relation $\Delta_{2p}$ with the horizontal
  relation $\Delta_{2n}$ below applied to all isotopes with $A > 30$.
  The colour scale is in keV and extrapolated isotopes are in
  black. \label{Fig. lige}}
\end{figure*}

\begin{figure*}
\centering
	\includegraphics{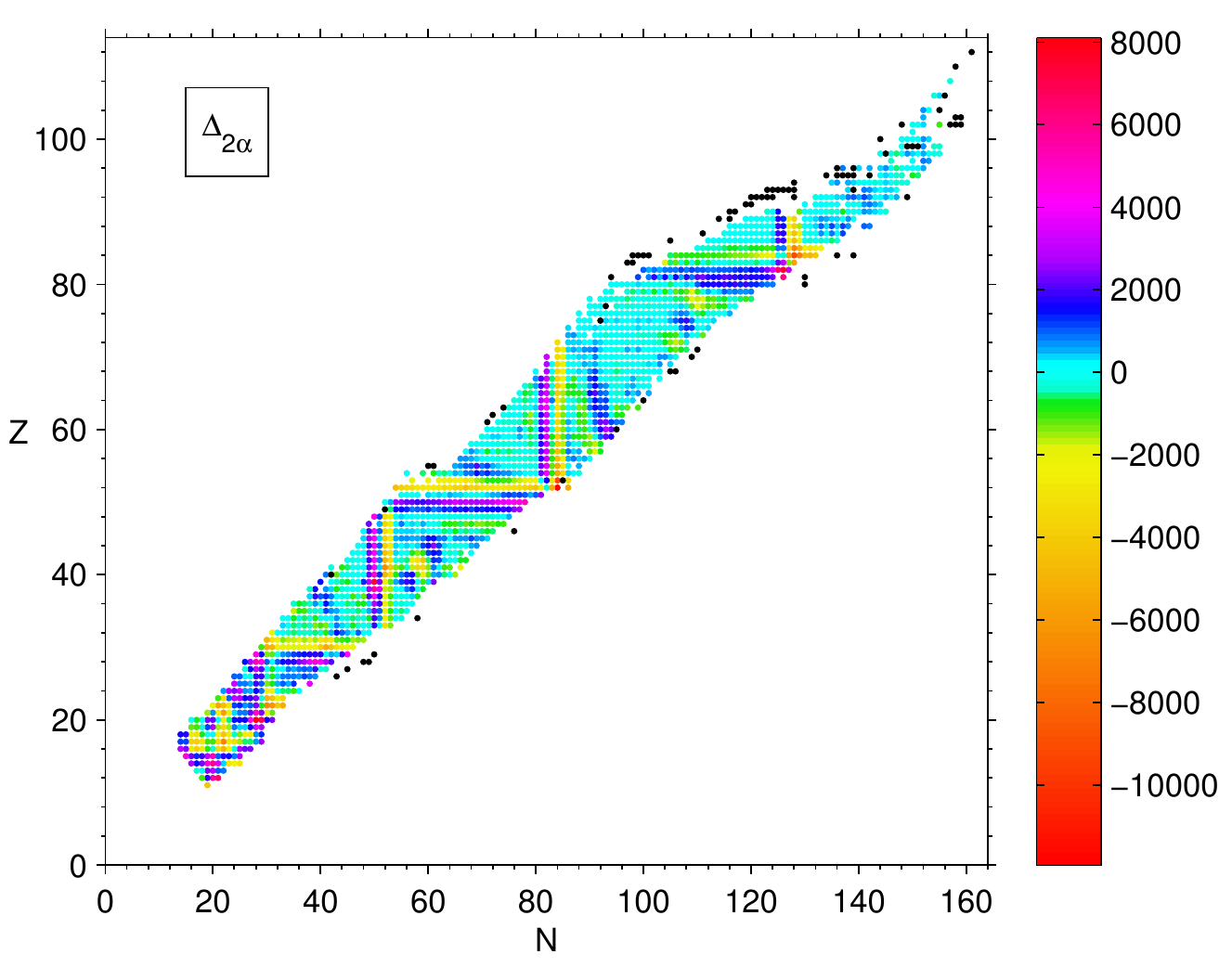}
	\includegraphics{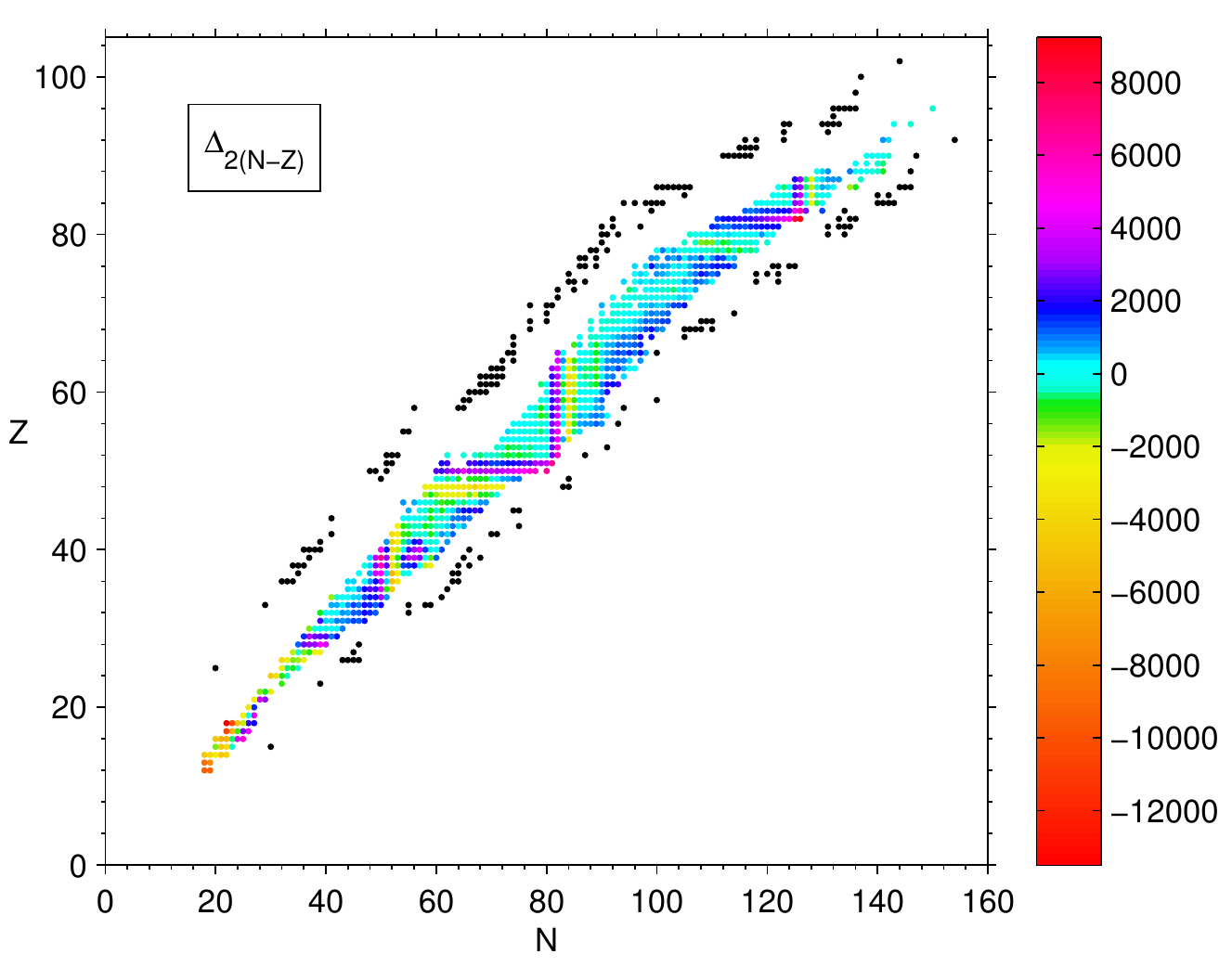}
\caption{The two diagonal relations, $\Delta_{2\alpha}$ above and
  $\Delta_{2(N-Z)}$ below applied to all isotopes with $A > 30$.  The
  colour scale is in keV and extrapolated isotopes are in
  black. \label{Fig. Dia}}
\end{figure*}

We now proceed to investigate the systematic behaviour of the mass
relations. The results from all four mass relations are shown in
figs.~\ref{Fig. lige} and \ref{Fig. Dia}. The most prominent
visible features arise from the shell effects around the more or less
magic numbers. Whenever a shell crossing is involved, a significant
deviation from the surrounding binding energies appear. How the mass
relation is positioned relative to the shell defines both the sign and
the scale of this deviation. 

Consequently, because the binding is amplified by the $+3$ coefficient
on the $(N,Z)$ value (see fig.~\ref{Structures.pdf}), a mass relation
computed for a magic number $(N,Z)$ must be significantly greater than
for neighbouring isotopes. A mass relation centred one or possibly
even two nucleons before a shell should also have a noticeably greater
outcome, though not to the same extent. Similarly, a relation centred
two or three nucleons after a shell would have a noticeably smaller
outcome, since the $-3$ coefficient in the relation would be closer to
the shell and would therefore dominate over the $+3$ contribution. If
the relation was centred just after a magic number the $-3$ and $+3$
coefficients would probably cancel, and the result might appear as if
unaffected by shells altogether.

Generally, it is tempting to assume that the extrapolations will be
more exact in regions with heavier isotopes, where changes from
isotope to isotope are more gradual. If the changes are more gradual,
the expected outcome should presumably be more reliable, as the
binding energies themselves would fluctuate less. This can also be
reflected in the attached uncertainties.

\subsection{Results from individual mass relations \label{Sect. Indi2}}

The $\Delta_{2n}$ and $\Delta_{2p}$ relations are shown in
fig.~\ref{Fig. lige}. They only combine nuclear masses
horizontally and vertically in the $N-Z$ diagram. They are therefore
well suited for extrapolations beyond neutron and proton drip lines,
but less well suited for the narrow strips of super-heavy elements.
These two mass relations are also only sensitive to their own type of
shell effects as seen in the figures. This confirms again the almost
independence of neutron and proton shell fillings. The very light
isotopes have been omitted, because they disrupted the energy scale
and made minor energy changes less obvious. Their binding energies
and structure are in any case strongly varying and any meaningful
extrapolation would be close to impossible.

The most prominent features in the $\Delta_{2p}$ relation on
fig.~\ref{Fig. lige} are the shells at $Z = 50$, and $82$, but also
the shell at $Z = 28$ is clearly visible. The trace of these shells
extend over numbers corresponding to the range of the mass
relations. As expected the influence is positive below and negative
above the shells. This symmetry extends to both sides of a shell, and
is reflected in the size as well. The absolute values at the shells
vary, but is always greater than $1 \, \si{\mega\electronvolt}$ and
often $\sim 2-4 \, \si{\mega\electronvolt}$. The results for $(N,Z)$
and $(N,Z-2)$, when located at the shell, are nearly identical with
opposite sign, which again demonstrates the symmetry of the shell effect.

It is also interesting to note how neutron shells only are visible with
$\Delta_{2p}$ at a proton shell, otherwise the mass relation is very
small. This emphasizes how exclusively $\Delta_{2p}$ is
concerned with effects relating to protons. The neutron and proton
shells are to a large extent, away from drip lines, filled
independently. The region around $Z = 40$ where $N > 50$, shows
many characteristics otherwise found in shells. There is an increase
in energy just before $Z = 40$ and a decrease in energy afterwards,
with a slight fluctuation at $Z = 40$, which is similar to the shell
at $Z = 82$. The energy changes are less pronounced than for other
shells and the energy changes are also less well-defined.
Nevertheless, the general smooth behaviour in the region is clearly
disrupted, and the result is compatible with $Z = 40$ as the most
prominent subshell.

Overall, the $\Delta_{2p}$ relation has away from shells, a very
pronounced tendency to more or less vanish. In particular, the region
beyond the $Z = 50$ shell is smooth and typically less than $500 \,
\si{\kilo\electronvolt}$ numerically. Extrapolations from this region should then
be very reliable. This claim will be carefully investigated in
sect.\ (\ref{Sec. Com. Res.}), where we also compare to extrapolations
from other mass relations.

The results in fig.~\ref{Fig. lige} from $\Delta_{2n}$ are
incredibly similar in most regards to the results for $\Delta_{2p}$.
The same tendency to complete cancellation is observed, though the
remains are typically less than $300 \, \si{\kilo\electronvolt}$ when
evaluated numerically. Actually, every visible feature appears more distinctly. The shells at $N = 28, 50, 82$, and $126$ are
not only obvious, they are sharply defined and confined to the area
immediately surrounding the shells. The symmetry around the shell
itself is also still present, and it is as clear as for protons. The
size of the shell deviations are $\sim 2-4 \,
\si{\mega\electronvolt}$, again very much comparable to the
$\Delta_{2p}$ results.

More interesting is the region around $(Z,N)=(40,60)$, where once
again a deviation is visible. The same region where the $Z = 40$
subshell was visible with $\Delta_{2p}$ now shows a deviation with
$\Delta_{2n}$. This is particularly interesting considering that none
of the major proton shells are visible away from a neutron shell,
which suggest that this is not solely a shell effect. The deviation
has some similarities with the other shells, but it is still
decisively different from an ordinary shell. Most strikingly, the
energy first increases, then decreases, and then increases again,
which again suggests that this effect arises from a more complicated
structure than a regular shell effect.

As an example of the possible use of these mass relations we look
into this mass region in a little more details. The
figs.~\ref{Fig. lige}a and \ref{Fig. lige}b clearly show shell
structures around $(Z,N) = (40,58)$.  First, the neutron shell at
$N=58$ is less prominent than the well established major shells, but
nevertheless unmistakingly recognized by the mass relations deviating
from zero.  This observation of a neutron subshell for $N=58,60$ is
discussed in \cite{hey11}. 

These shells for $Z=40$ and $N=58$ do not extend through all the
known isotopes. For $Z=40$, the structure is absent for $N<49$, and
present for $49< N<63 $. For $N=58$, the structure is absent for
$Z>42$, and present for $38<Z<42$. The explanations can be found by
inspecting the fillings of the corresponding neutron and proton
shells.  For nucleon numbers between $40$ and $50$, the $g_{9/2}$
shell is only partly occupied. Adding more nucleons require
occupation of other shells, that is $g_{7/2}, d_{5/2}$ and possibly
$s_{1/2}$.  Then the neutron shell at $N=58$ disappears when $Z$
increases beyond $42$.  This is precisely when at least 4 protons
occupy the $g_{9/2}$ shell which therefore wants to deform to avoid
the degeneracy.  The neutron shell is not sufficiently strong to
prevent this deformation.  For $Z=40$, the proton shell is only
visible for $N$ larger than $48$, which is when the rather close-lying
$g_{7/2}$ and $ d_{5/2}$ levels begin to be occupied.  The gain in
neutron deformation energy is not sufficient to overcome the rather
strong spherical proton shell effect.  The reason is that the neutron
single-particle level density only changes relatively little with
modest deformation.

The results from using the diagonal relations for $\Delta_{2\alpha}$
and $\Delta_{2(N-Z)}$ are presented in fig.~\ref{Fig. Dia}. The
$\Delta_{2\alpha}$ relation is oriented diagonally towards the heavy
isotopes in the chart of nuclides. It should therefore be able to
extrapolate to heavier isotopes than any of the other mass relation.
Unfortunately, this orientation also rather strongly confines it to
the isotopes at the heavy end of known isotopes. 

The shells are again very pronounced, but now all the structures from
both $\Delta_{2p}$ and $\Delta_{2n}$ appear in $\Delta_{2\alpha}$. It
is interesting to notice how the neutron shells are more sharply
defined than the proton shells, as it also appeared when comparing the
results of $\Delta_{2p}$ and $\Delta_{2n}$. Not surprisingly, the
deviation around $(Z,N)=(40,60)$ is even more prominent here, but also the
area around $(Z,N)=(60,92)$ shows a rather strong deviation from zero.
This deviation could also be detected with $\Delta_{2n}$, albeit more
faintly, but it was almost invisible with $\Delta_{2p}$. A clear and significant effect in this mass region is therefore
somewhat surprising, but it demonstrates how well $\Delta_{2\alpha}$
detects the more elusive tendencies.

As $\Delta_{2\alpha}$ includes more effects than $\Delta_{2p}$ and
$\Delta_{2n}$, the results also vary much more. The reliability of
any extrapolated result might therefore be questionable. This objection
is legitimate for extrapolations involving several different shells.
However, away from shells the fluctuations around zero are generally
less than $500\, \si{\kilo\electronvolt}$. In particular, a promising
mass region with smooth behaviour is $N > 126$ and $Z > 82$. For heavy
or super-heavy isotopes the results should be as reliable as with
$\Delta_{2p}$ or $\Delta_{2n}$ at the drip lines. This suggests
interesting extrapolations with $\Delta_{2\alpha}$ in the less
accessible region of heavy or super-heavy nuclei.  

Finally, the other diagonal relation, $\Delta_{2(N-Z)}$, is shown in
fig.\ (\ref{Fig. Dia}). Unfortunately, it is oriented perpendicular
to the rather narrow strip of measured masses. The number of isotopes
to which this relation can be readily applied is therefore rather
limited. On the other hand it points directly towards the boundary of
the known nuclear territory which then should allow extrapolations coinciding with $\Delta_{2n}$ and $\Delta_{2p}$.
However, the general behaviour is the same as with $\Delta_{2\alpha}$,
and both proton and neutron shells are clearly visible. This limits
the possibilities for reliable extrapolations. 

These discussions suggest that the mass relation with the largest
extrapolation potential seem to be the $\Delta_{2n}$ relation. It
generally cancels completely, is affected by few unpredictable
effects, and it sharply defines the neutron shells. The same is true
for the $\Delta_{2p}$ relation, although its nature is slightly more
erratic. The diagonal relations $\Delta_{2(N-Z)}$ and
$\Delta_{2\alpha}$ should be as reliable as $\Delta_{2n}$ or
$\Delta_{2p}$, but more care must be taken when applying them, as they
are more often influenced by shells.

Viewed collectively the results unanimously corroborate the
predictions from sect.\ (\ref{Sect. mass rel}), which in turn
indicates that the initial division of the binding energy in 
the three characteristic terms is well founded.

\subsection{Evaluations with $\mathbf{Q_{\alpha}}$ values \label{Sec. Q}}

The available measurements of $Q_{\alpha}$ values extend to far
heavier isotopes than the binding
energies, although recent developments at SHIPTRAP \cite{Minaya} in measuring absolute masses may allow for more extensive use of
our method in the future. $Q_{\alpha}$ can therefore
provide greater insight into the nature of the super-heavy isotopes.
Applying $Q_{\alpha}$ values as outlined in eq.\ (\ref{Qalph}) have a
number of advantages. Not only are they measured to a higher nucleon
number, but eq.\ (\ref{Qalph}) only employs three different
measurements, which gives a more compact relation, more likely to be
applicable. Unfortunately, the measured $Q_{\alpha}$ values do not necessarily
relate to ground state configurations. The specific state, in
particular, among the super-heavy isotopes are usually unknown. This
fact alone makes it very difficult to extrapolate binding energies
accurately from chains of connected $Q_{\alpha}$ measurements. As a
consequence we shall only use the results to shed light on the general
tendencies of the binding energies in the super-heavy region.

\begin{figure*}
\centering
	\includegraphics{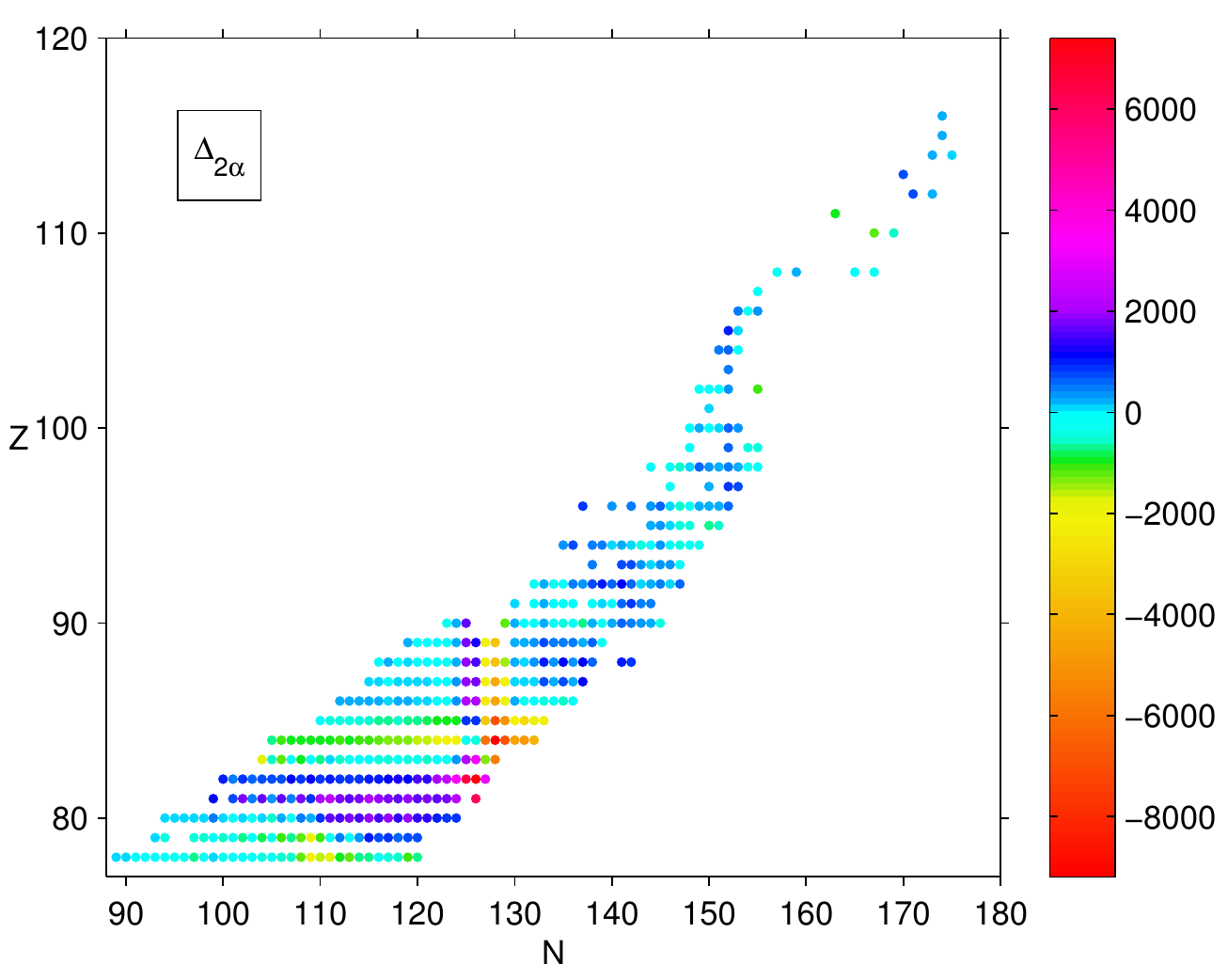}
	\includegraphics{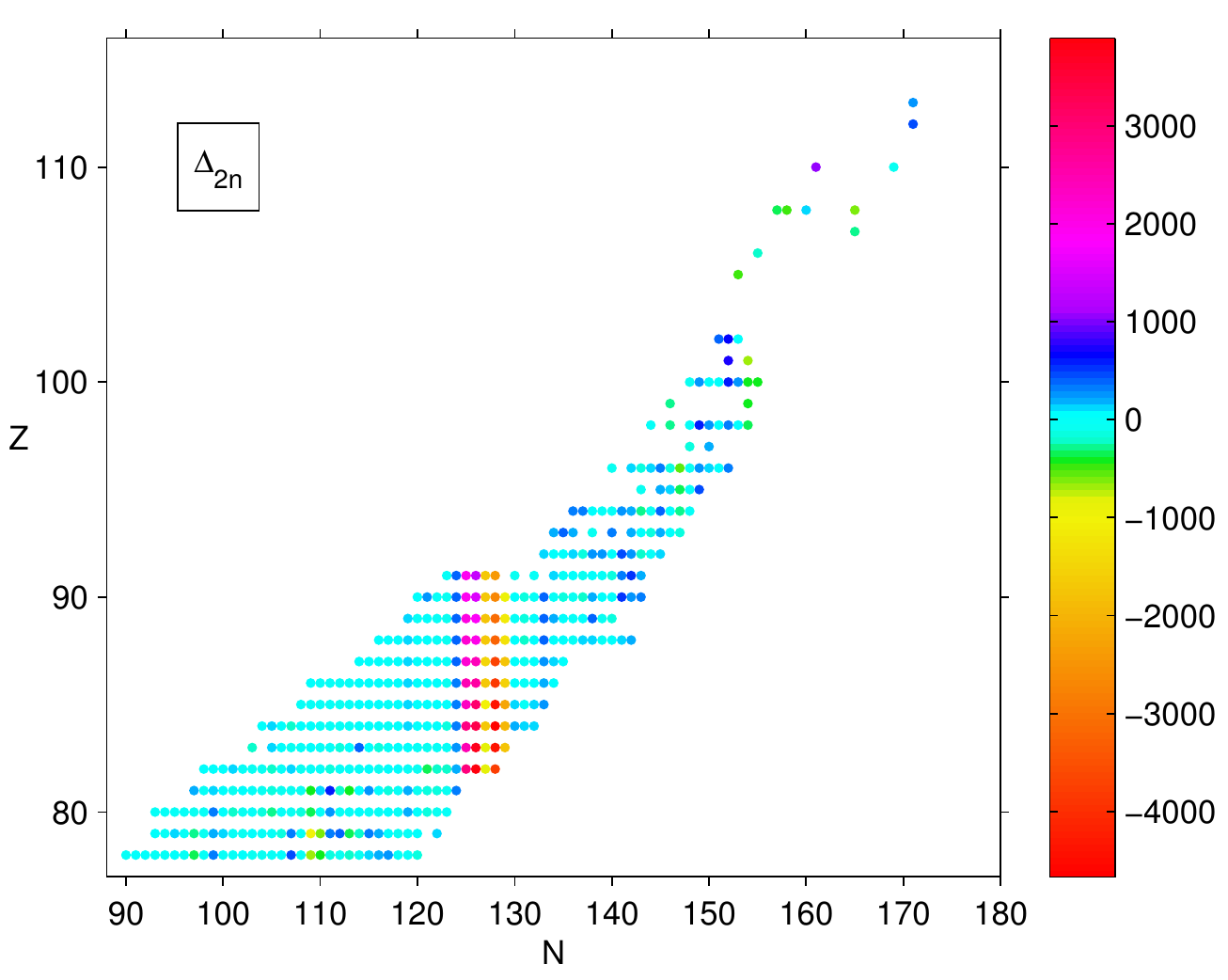}
\caption{Results of evaluating $Q_{\alpha}$ values with the $\Delta_{2\alpha}$ relation above and the $\Delta_{2n}$ relation below. The evaluations are confined to the superheavy isotopes. The colour scale is still in keV. \label{Fig. Qa}}
\end{figure*}

The results for the super-heavy region are displayed in
fig.~\ref{Fig. Qa} for both $\Delta_{2\alpha}$ and $\Delta_{2n}$
relations. We notice first of all a rather clear picture of the
deviation from zero around the known high-end shells at $N=126$ and
$Z=82$. The size of the deviations is $\sim 2-4 \,
\si{\mega\electronvolt}$ both positive and negative. 

Otherwise, the most interesting feature in fig.~\ref{Fig. Qa}, visible
with both $\Delta_{2\alpha}$ and $\Delta_{2n}$, is the systematic
non-zero values around $N = 152$. This deviation extend through all
evaluations with $\Delta_{2\alpha}$, and with $\Delta_{2n}$ significant
positive deviations are also visible. This is exactly the behaviour
expected from a minor shell, and with $\Delta_{2n}$ the deviations are
even symmetric around the shell. The size of the deviations are $\sim
1 \, \si{\mega\electronvolt}$, so it is rather weak compared with
other shells. This is not surprising considering the nucleon number it
occurs at, but these features definitely corresponds to that of an
ordinary shell. On the other hand, no other shell effect appears even
though the continued increasing stability demonstrate that some shell
effects provide the necessary smaller binding energy.

\section{Extrapolated binding energies \label{Sec. Com. Res.}}

We shall compare the results from different extrapolations, and define
a suitable average leading to better accuracy. Any systematic
discrepancies between the individual relations should be
accentuated by such a combination. Particularly interesting are applications
in the region where $Z > 82$ and $N < 126$,
which is less known. The estimations presented by Audi and Meng
\cite{Audi} for instance diligently cover most of the chart of
nuclides except for this specific area. After the general discussion
we shall provide tables of extrapolated nuclear binding energies.

\subsection{Improving the accuracy }

The individual extrapolations are all legitimate attempts at
estimating the binding energy of unknown isotopes. In the landscape of
binding energies each linear mass relation can be seen as approaching
the unknown isotope from a different direction. A single mass relation
cannot be expected to provide perfect predictions, because of the
fluctuating, possibly chaotic, nature of the binding energy. However,
if a particular isotope could be approached from several different
directions, the expected fluctuations could be viewed from several
sides, which would provide a clearer image of the given isotope. In
other words, if an isotope could be extrapolated by several different
mass relations, the results could be examined and used either to
select the most accurate of the extrapolations or combined to provide a
much more reliable estimate of the binding energy.

Comparing different extrapolations would also examine the legitimacy
of the method itself. If different extrapolations for the same isotope
deviated significantly it could cast doubt on the entire
procedure. Different mass relations will, of course, involve shells or
other influencing factors at different isotopes, and care must be
taken when comparing the extrapolations.

There are several considerations to bear in mind, when combining
extrapolations based on different mass relations. Both when selecting
or prioritizing particular extrapolations, and when calculating an
appropriate uncertainty for the final result. To avoid confusion the
exact procedure leading to the recommended results will first be
explained in some detail.

First of all, it is vital that the individual extrapolations seem
reliable, and have a certain degree of precision. To accommodate these
requirements only extrapolations with a limited uncertainty,
specifically extrapolations where $\sigma_i < 500 \,
\si{\kilo\electronvolt}$, are included in the calculated average.
However, to demonstrate the possible fallibility of the individual
mass relations in certain areas, all extrapolations of relevant
isotopes, even if unused, are included in
table \ref{Tab. Ex}. The final column in table \ref{Tab. Ex}
lists the mass relations used in the calculated average.

The actual weighted average is calculated in stages to both account
for known effects such as shells and to detect general deviations.
First we use all extrapolations with $\sigma_i < 500 \,
\si{\kilo\electronvolt}$ to provide $r_i \pm \sigma_i$. Second, we define
the relation
\begin{align}
f_i = \frac{|r_i - r|}{\sigma_i} \label{f_i}  \;,
\end{align}
where $r$ denotes the average, and $r_i$ is the individual
extrapolation. If $\max(f_i) < 3/2$, then the individual
extrapolations are within an acceptable range of the average, and the
uncertainty for the average value is defined as $\sigma =
\min(\sigma_i)$.

Otherwise, if $\max(f_i) > 3/2$ the individual extrapolations differ
too greatly from the average result, and the computed average has little
meaning. Any extrapolations involving shell crossings or crossing the
$N=Z$-line often differs from the general tendencies, as apparent from
figs.~\ref{Fig. lige} and \ref{Fig. Dia}. These results are marked
with a * symbol in table~\ref{Tab. Ex}, and are considered less
reliable. The marked extrapolations are then excluded, and a new, more
plausible, average is calculated. Based on this average a relation
similar to the one presented in eq.\ (\ref{f_i}) is defined, though
now obtained by fewer extrapolations. Once again, if $\max(f_i) < 3/2$,
then the uncertainty is defined as $\sigma = \min(\sigma_i)$.

The extrapolations may be incompatible as in the neighbourhood of
closed shells or by crossing the $N=Z$ line. Then a meaningful
uncertainty is defined as $\sigma = 2/3 \, \max(|r_i - r|)$, where $r_i$ only
includes the extrapolations used in the final calculation of $r$. As
a consequence of this procedure some results might be based on a
single extrapolation, even though multiple mass relations have
estimated the isotope. The other available extrapolations could for instance involve a shell crossing, and would then be discarded if the initial results were incompatible. Actually, some isotopes, which have been
estimated by several mass relations, might not have a meaningful
resulting average at all, if all the extrapolations had an 
uncertainty exceeding 500 keV. The isotopes in both cases, those with either an average based on a single extrapolation or no average at all, have been
omitted from table \ref{Tab. Ex}, as they provided no relevant
information.

\newpage

\oddsidemargin=-1.5cm 

\LTcapwidth=0.8\textwidth 

\begin{longtable*}{p{0.08\textwidth} *{4}{D{.}{.}{8.1} @{$\pm$} D{.}{.}{4.1}} *{1}{D{.}{.}{8.1} @{$\pm$} D{.}{.}{4.2}} p{0.21\textwidth}} 

\colrule
Nucleus & \multicolumn{8}{c}{Estimates of binding energies [keV]}  &  \multicolumn{2}{c}{Average [keV]} & Applied Relations\\ 
 \cline{2-9} 
(Z, N)   &   \multicolumn{2}{c}{$\Delta_{2n}$}  & \multicolumn{2}{c}{$\Delta_{2p}$} & \multicolumn{2}{c}{$\Delta_{2\alpha}$}  & \multicolumn{2}{c}{$\Delta_{2(N-Z)}$}  &     \multicolumn{2}{c}{} &         \\

\colrule
\endhead

\colrule
\multicolumn{12}{c}{Continued on next page} \\ 
\colrule
\endfoot

\colrule \\
\caption{The results of combining the
  extrapolations based on the four different mass relations. The final
  column indicates which relations were used when calculating a
  particular average. Only extrapolations with an uncertainty
  $\sigma_i < 500 \, \si{\kilo\electronvolt}$ were considered when
  calculating this average. If an extrapolation includes isotopes
  influence by either a shell or on the $N = Z$ line, it is marked
  with a * symbol. \label{Tab. Ex}} \endlastfoot

( 26, 43) & 576250 &1292 &* 577468 &239 &* 577029 &202 &* 578091 &275 &577422 &446 & $\Delta_{2p}$,\, $\Delta_{2\alpha}$,\, $\Delta_{2(N-Z)}$\\
( 26, 44) & 579206 &2256 &* 582840 &250 &    \multicolumn{2}{c}{-} &* 584436 &276 &583561 &583 & $\Delta_{2p}$,\, $\Delta_{2(N-Z)}$\\
( 27, 45) & 597521 &2633 &* 601637 &257 &* 601272 &152 &* 601465 &361 &601378 &152 & $\Delta_{2p}$,\, $\Delta_{2\alpha}$,\, $\Delta_{2(N-Z)}$\\
( 28, 46) & 623965 &648 &* 623861 &297 &* 619868 &2212 &* 624385 &203 &624218 &203 & $\Delta_{2p}$,\, $\Delta_{2(N-Z)}$\\
( 32, 55) &* 719433 &237 & 721976 &1353 &    \multicolumn{2}{c}{-} &* 718789 &410 &719272 &237 &$\Delta_{2n}$,\, $\Delta_{2(N-Z)}$\\
( 33, 55) &* 733407 &256 & 735628 &1530 & 738781 &2351 &* 732688 &394 &733193 &256 &$\Delta_{2n}$,\, $\Delta_{2(N-Z)}$\\
( 36, 33) & 561880 &486 & 561002 &1513 &    \multicolumn{2}{c}{-} & 565048 &294 &564201 &1548 &$\Delta_{2n}$,\, $\Delta_{2(N-Z)}$\\
( 36, 34) &* 580478 &192 &* 582467 &2134 &    \multicolumn{2}{c}{-} & 580221 &276 &580395 &192 &$\Delta_{2n}$,\, $\Delta_{2(N-Z)}$\\
( 36, 62) & 806662 &366 & 807701 &185 &    \multicolumn{2}{c}{-} & 805037 &1328 &807490 &552 &$\Delta_{2n}$,\, $\Delta_{2p}$\\
( 36, 63) & 808943 &422 & 811736 &540 &    \multicolumn{2}{c}{-} & 811516 &78 &811431 &1659 &$\Delta_{2n}$,\, $\Delta_{2(N-Z)}$\\
( 37, 35) &* 592557 &489 &* 594706 &1896 &    \multicolumn{2}{c}{-} & 593650 &362 &593264 &362 &$\Delta_{2n}$,\, $\Delta_{2(N-Z)}$\\
( 38, 64) & 846621 &1136 & 845897 &155 &    \multicolumn{2}{c}{-} & 846966 &256 &846184 &521 & $\Delta_{2p}$,\, $\Delta_{2(N-Z)}$\\
( 40, 38) &    \multicolumn{2}{c}{-} &* 642144 &166 &    \multicolumn{2}{c}{-} & 642129 &151 &642136 &151 & $\Delta_{2p}$,\, $\Delta_{2(N-Z)}$\\
( 40, 42) & 694171 &4774 & 694590 &230 & 694624 &145 & 696013 &1606 &694614 &145 & $\Delta_{2p}$,\, $\Delta_{2\alpha}$\\
( 42, 70) & 928505 &382 & 928726 &2999 &    \multicolumn{2}{c}{-} & 929351 &271 &929068 &271 &$\Delta_{2n}$,\, $\Delta_{2(N-Z)}$\\
( 45, 74) & 988066 &193 &* 986373 &1081 &* 986234 &1019 &* 987801 &216 &987949 &193 &$\Delta_{2n}$,\, $\Delta_{2(N-Z)}$\\
( 45, 75) & 991514 &260 &* 990420 &1214 &* 991092 &2681 &* 991504 &176 &991507 &176 &$\Delta_{2n}$,\, $\Delta_{2(N-Z)}$\\
( 48, 48) &* 797281 &190 &* 793119 &175 &    \multicolumn{2}{c}{-} &* 795277 &1121 &795033 &1499 &$\Delta_{2n}$,\, $\Delta_{2p}$\\
( 48, 83) &    \multicolumn{2}{c}{-} &*1080376 &381 &    \multicolumn{2}{c}{-} &*1084581 &197 &1083690 &2209 & $\Delta_{2p}$,\, $\Delta_{2(N-Z)}$\\
( 49, 84) &*1097714 &232 &*1095129 &400 &    \multicolumn{2}{c}{-} &*1102390 &87 &1101543 &4276 &$\Delta_{2n}$,\, $\Delta_{2p}$,\, $\Delta_{2(N-Z)}$\\
( 50, 51) &* 835352 &287 &* 836267 &295 &    \multicolumn{2}{c}{-} &* 835137 &117 &835299 &646 &$\Delta_{2n}$,\, $\Delta_{2p}$,\, $\Delta_{2(N-Z)}$\\
( 52, 52) &* 848264 &323 &* 852185 &353 &    \multicolumn{2}{c}{-} &* 850904 &194 &850562 &1532 &$\Delta_{2n}$,\, $\Delta_{2p}$,\, $\Delta_{2(N-Z)}$\\
( 52, 87) &*1138134 &1243 &1140623 &270 &    \multicolumn{2}{c}{-} &*1140110 &265 &1140362 &265 & $\Delta_{2p}$,\, $\Delta_{2(N-Z)}$\\
( 53, 85) &*1144407 &1509 &1143992 &221 &1144890 &266 &*1142366 &2237 &1144358 &355 & $\Delta_{2p}$,\, $\Delta_{2\alpha}$\\
( 58, 94) &1240393 &202 &1241112 &409 &    \multicolumn{2}{c}{-} &1242205 &155 &1241501 &738 &$\Delta_{2n}$,\, $\Delta_{2p}$,\, $\Delta_{2(N-Z)}$\\
( 60, 68) &1047404 &187 &1049530 &2129 &    \multicolumn{2}{c}{-} &1046496 &93 &1046674 &487 &$\Delta_{2n}$,\, $\Delta_{2(N-Z)}$\\
( 60, 95) &1268120 &682 &1266702 &277 &1266251 &423 &1266860 &771 &1266566 &277 & $\Delta_{2p}$,\, $\Delta_{2\alpha}$\\
( 61, 70) &1070175 &395 &1070384 &762 &    \multicolumn{2}{c}{-} &1068999 &333 &1069488 &458 &$\Delta_{2n}$,\, $\Delta_{2(N-Z)}$\\
( 61, 71) &1079011 &1534 &1080538 &1146 &1079524 &388 &1079127 &303 &1079277 &303 & $\Delta_{2\alpha}$,\, $\Delta_{2(N-Z)}$\\
( 62, 70) &    \multicolumn{2}{c}{-} &1072374 &446 &    \multicolumn{2}{c}{-} &1072246 &122 &1072254 &122 & $\Delta_{2p}$,\, $\Delta_{2(N-Z)}$\\
( 62, 71) &1081835 &1800 &1081867 &393 &    \multicolumn{2}{c}{-} &1082341 &181 &1082257 &181 & $\Delta_{2p}$,\, $\Delta_{2(N-Z)}$\\
( 62, 72) &1094145 &2024 &1094109 &362 &1094870 &209 &1094296 &165 &1094471 &266 & $\Delta_{2p}$,\, $\Delta_{2\alpha}$,\, $\Delta_{2(N-Z)}$\\
( 62, 98) &1302650 &448 &1302344 &214 &    \multicolumn{2}{c}{-} &1302713 &1063 &1302401 &214 &$\Delta_{2n}$,\, $\Delta_{2p}$\\
( 63, 72) &    \multicolumn{2}{c}{-} &1094922 &283 &    \multicolumn{2}{c}{-} &1093656 &499 &1094614 &638 & $\Delta_{2p}$,\, $\Delta_{2(N-Z)}$\\
( 63, 74) &1117498 &3591 &1116361 &210 &1117409 &344 &1115580 &695 &1116645 &509 &$\Delta_{2p}$,\, $\Delta_{2\alpha}$\\
( 64, 74) &1120095 &3568 &1119918 &109 &    \multicolumn{2}{c}{-} &1119650 &144 &1119820 &109 & $\Delta_{2p}$,\, $\Delta_{2(N-Z)}$\\
( 64,100) &1333063 &248 &1332671 &328 &1333134 &489 &1334078 &641 &1332950 &248 &$\Delta_{2n}$,\, $\Delta_{2p}$,\, $\Delta_{2\alpha}$\\
( 64,101) &1337977 &453 &1337702 &295 &    \multicolumn{2}{c}{-} &1338532 &730 &1337784 &295 &$\Delta_{2n}$,\, $\Delta_{2p}$\\
( 65,100) &1341837 &190 &1341306 &491 &1339899 &757 &1342848 &424 &1341928 &613 &$\Delta_{2n}$,\, $\Delta_{2p}$,\, $\Delta_{2(N-Z)}$\\
( 67,105) &1387302 &333 &1387183 &498 &1385586 &1250 &1388720 &385 &1387762 &639 &$\Delta_{2n}$,\, $\Delta_{2p}$,\, $\Delta_{2(N-Z)}$\\
( 68,105) &1396784 &436 &1398240 &323 &1397201 &58 &1398343 &161 &1397349 &663 &$\Delta_{2n}$,\, $\Delta_{2p}$,\, $\Delta_{2\alpha}$,\, $\Delta_{2(N-Z)}$\\
( 68,106) &1403282 &249 &1404464 &609 &1404991 &357 &1404826 &129 &1404544 &841 &$\Delta_{2n}$,\, $\Delta_{2\alpha}$,\, $\Delta_{2(N-Z)}$\\
( 69,108) &1422022 &184 &1423137 &998 &    \multicolumn{2}{c}{-} &1422893 &93 &1422716 &463 &$\Delta_{2n}$,\, $\Delta_{2(N-Z)}$\\
( 69,109) &1426615 &361 &1428561 &1044 &    \multicolumn{2}{c}{-} &1427412 &474 &1426908 &361 &$\Delta_{2n}$,\, $\Delta_{2(N-Z)}$\\
( 70,109) &1436405 &360 &1435954 &1439 &1436501 &103 &1435908 &635 &1436494 &103 &$\Delta_{2n}$,\, $\Delta_{2\alpha}$\\
( 71, 83) &*1227402 &438 &1227271 &471 &    \multicolumn{2}{c}{-} &*1226192 &779 &1227341 &438 &$\Delta_{2n}$,\, $\Delta_{2p}$\\
( 71,110) &1450595 &425 &1448160 &1502 &1450074 &93 &1450005 &593 &1450097 &93 &$\Delta_{2n}$,\, $\Delta_{2\alpha}$\\
( 72, 85) &1249894 &187 &1249664 &319 &    \multicolumn{2}{c}{-} &1250965 &638 &1249835 &187 &$\Delta_{2n}$,\, $\Delta_{2p}$\\
( 73, 85) &1250003 &404 &1249843 &497 &    \multicolumn{2}{c}{-} &1250054 &235 &1250013 &235 &$\Delta_{2n}$,\, $\Delta_{2p}$,\, $\Delta_{2(N-Z)}$\\
( 74, 87) &1271494 &365 &1272211 &113 &    \multicolumn{2}{c}{-} &1272571 &361 &1272183 &459 &$\Delta_{2n}$,\, $\Delta_{2p}$,\, $\Delta_{2(N-Z)}$\\
( 75, 87) &1271867 &492 &1271718 &302 &    \multicolumn{2}{c}{-} &1270835 &507 &1271759 &302 &$\Delta_{2n}$,\, $\Delta_{2p}$\\
( 75, 92) &1324297 &207 &1324669 &311 &1323990 &403 &1325523 &1207 &1324346 &207 &$\Delta_{2n}$,\, $\Delta_{2p}$,\, $\Delta_{2\alpha}$\\
( 75,118) &1529546 &345 &*1529841 &1167 &*1530772 &2269 &*1530133 &429 &1529776 &345 &$\Delta_{2n}$,\, $\Delta_{2(N-Z)}$\\
( 76, 89) &1293730 &390 &1293974 &229 &    \multicolumn{2}{c}{-} &1294570 &340 &1294078 &229 &$\Delta_{2n}$,\, $\Delta_{2p}$,\, $\Delta_{2(N-Z)}$\\
( 77, 88) &1282791 &307 &1283328 &429 &    \multicolumn{2}{c}{-} &1281841 &531 &1282973 &307 &$\Delta_{2n}$,\, $\Delta_{2p}$\\
( 77, 93) &1335380 &180 &1335618 &220 &1335199 &400 &1336152 &643 &1335446 &180 &$\Delta_{2n}$,\, $\Delta_{2p}$,\, $\Delta_{2\alpha}$\\
( 78, 91) &1314656 &761 &1315246 &242 &    \multicolumn{2}{c}{-} &1314896 &292 &1315104 &242 & $\Delta_{2p}$,\, $\Delta_{2(N-Z)}$\\
( 79, 90) &1303813 &240 &1304075 &282 &    \multicolumn{2}{c}{-} &1303200 &380 &1303787 &391 &$\Delta_{2n}$,\, $\Delta_{2p}$,\, $\Delta_{2(N-Z)}$\\
( 79, 95) &1356317 &297 &1356279 &181 &    \multicolumn{2}{c}{-} &1355432 &528 &1356290 &181 &$\Delta_{2n}$,\, $\Delta_{2p}$\\
( 80, 93) &1335989 &479 &1336045 &258 &    \multicolumn{2}{c}{-} &1336382 &448 &1336104 &258 &$\Delta_{2n}$,\, $\Delta_{2p}$,\, $\Delta_{2(N-Z)}$\\
( 80,131) &    \multicolumn{2}{c}{-} &*1644785 &366 &    \multicolumn{2}{c}{-} &*1643683 &294 &1644115 &447 & $\Delta_{2p}$,\, $\Delta_{2(N-Z)}$\\
( 81, 97) &1377933 &406 &*1377553 &446 &    \multicolumn{2}{c}{-} &*1377229 &220 &1377415 &220 &$\Delta_{2n}$,\, $\Delta_{2p}$,\, $\Delta_{2(N-Z)}$\\
( 82, 94) &1346745 &126 &*1347085 &243 &*1346991 &1303 &*1347653 &636 &1346817 &126 &$\Delta_{2n}$,\, $\Delta_{2p}$\\
( 82, 95) &1356385 &434 &*1356892 &452 &    \multicolumn{2}{c}{-} &*1357465 &853 &1356628 &434 &$\Delta_{2n}$,\, $\Delta_{2p}$\\
( 83, 98) &    \multicolumn{2}{c}{-} &*1389816 &321 &*1390047 &488 &*1391306 &1084 &1389886 &321 & $\Delta_{2p}$,\, $\Delta_{2\alpha}$\\
( 84, 96) &    \multicolumn{2}{c}{-} &*1366448 &262 &    \multicolumn{2}{c}{-} &*1366742 &156 &1366665 &156 & $\Delta_{2p}$,\, $\Delta_{2(N-Z)}$\\
( 84, 98) &    \multicolumn{2}{c}{-} &*1389136 &311 &*1388890 &132 &*1389155 &80 &1389087 &80 & $\Delta_{2p}$,\, $\Delta_{2\alpha}$,\, $\Delta_{2(N-Z)}$\\
( 84, 99) &    \multicolumn{2}{c}{-} &*1399474 &424 &*1398674 &463 &*1398201 &227 &1398514 &640 & $\Delta_{2p}$,\, $\Delta_{2\alpha}$,\, $\Delta_{2(N-Z)}$\\
( 84,140) &    \multicolumn{2}{c}{-} &1712129 &227 &    \multicolumn{2}{c}{-} &1711176 &253 &1711704 &352 & $\Delta_{2p}$,\, $\Delta_{2(N-Z)}$\\
( 84,141) &    \multicolumn{2}{c}{-} &1715816 &384 &    \multicolumn{2}{c}{-} &1714216 &141 &1714407 &939 & $\Delta_{2p}$,\, $\Delta_{2(N-Z)}$\\
( 84,142) &    \multicolumn{2}{c}{-} &1720838 &290 &    \multicolumn{2}{c}{-} &1719998 &55 &1720027 &541 & $\Delta_{2p}$,\, $\Delta_{2(N-Z)}$\\
( 84,143) &    \multicolumn{2}{c}{-} &1724348 &237 &    \multicolumn{2}{c}{-} &1723840 &270 &1724127 &237 & $\Delta_{2p}$,\, $\Delta_{2(N-Z)}$\\
( 86,103) &    \multicolumn{2}{c}{-} &*1438847 &349 &    \multicolumn{2}{c}{-} &*1438645 &271 &1438721 &271 & $\Delta_{2p}$,\, $\Delta_{2(N-Z)}$\\
( 86,144) &1747230 &93 &1747198 &173 &    \multicolumn{2}{c}{-} &1746714 &170 &1747127 &276 &$\Delta_{2n}$,\, $\Delta_{2p}$,\, $\Delta_{2(N-Z)}$\\
( 86,145) &1751216 &89 &    \multicolumn{2}{c}{-} &    \multicolumn{2}{c}{-} &1750244 &110 &1750834 &393 &$\Delta_{2n}$,\, $\Delta_{2(N-Z)}$\\
( 87,111) &1520866 &442 &*1519842 &1133 &*1520530 &312 &*1520839 &1472 &1520642 &312 &$\Delta_{2n}$,\, $\Delta_{2\alpha}$\\
( 88,146) &1772938 &147 &    \multicolumn{2}{c}{-} &    \multicolumn{2}{c}{-} &1771596 &496 &1772829 &822 &$\Delta_{2n}$,\, $\Delta_{2(N-Z)}$\\
( 90,116) &1572224 &841 &1572457 &1181 &1572136 &212 &1572279 &167 &1572224 &167 &$\Delta_{2\alpha}$,\, $\Delta_{2(N-Z)}$\\
( 90,117) &1580086 &1803 &1581275 &1135 &1580703 &450 &1580524 &489 &1580621 &450 &$\Delta_{2\alpha}$,\, $\Delta_{2(N-Z)}$\\
( 90,147) &1791915 &232 &    \multicolumn{2}{c}{-} &    \multicolumn{2}{c}{-} &1792436 &256 &1792150 &232 &$\Delta_{2n}$,\, $\Delta_{2(N-Z)}$\\
( 92,123) &    \multicolumn{2}{c}{-} &1641179 &2031 &1638388 &342 &*1639006 &358 &1638683 &342 &$\Delta_{2\alpha}$,\, $\Delta_{2(N-Z)}$\\
( 92,149) &1816700 &422 &    \multicolumn{2}{c}{-} &1816384 &247 &    \multicolumn{2}{c}{-} &1816465 &247 &$\Delta_{2n}$,\, $\Delta_{2\alpha}$\\
( 94,131) &    \multicolumn{2}{c}{-} &1706355 &424 &    \multicolumn{2}{c}{-} &1705431 &131 &1705512 &562 &$\Delta_{2p}$,\, $\Delta_{2(N-Z)}$\\
( 94,132) &1714752 &162 &1714640 &303 &    \multicolumn{2}{c}{-} &1714508 &50 &1714531 &50 &$\Delta_{2n}$,\, $\Delta_{2p}$,\, $\Delta_{2(N-Z)}$\\
( 94,154) &1856744 &85 &1856600 &359 &    \multicolumn{2}{c}{-} &    \multicolumn{2}{c}{-} &1856736 &85 &$\Delta_{2n}$,\, $\Delta_{2p}$\\
( 95,132) &    \multicolumn{2}{c}{-} &1715359 &480 &    \multicolumn{2}{c}{-} &1714860 &254 &1714969 &254 & $\Delta_{2p}$,\, $\Delta_{2(N-Z)}$\\
( 96,134) &    \multicolumn{2}{c}{-} &1734758 &458 &    \multicolumn{2}{c}{-} &1734312 &174 &1734368 &174 & $\Delta_{2p}$,\, $\Delta_{2(N-Z)}$\\
( 96,136) &1751675 &294 &    \multicolumn{2}{c}{-} &1752023 &382 &1750297 &415 &1751444 &765 &$\Delta_{2n}$,\, $\Delta_{2\alpha}$,\, $\Delta_{2(N-Z)}$\\
( 98,140) &1786792 &243 &1787021 &237 &1786900 &712 &    \multicolumn{2}{c}{-} &1786909 &237 &$\Delta_{2n}$,\, $\Delta_{2p}$\\
(100,144) &1822301 &819 &1821948 &172 &1822088 &327 &    \multicolumn{2}{c}{-} &1821979 &172 &$\Delta_{2p}$,\, $\Delta_{2\alpha}$\\

\end{longtable*}

\oddsidemargin=-0.5cm 

\subsection{Numerical results }

All the results presented in table~\ref{Tab. Ex} have merit and
provide some information, though not all will be commented
on. Instead, focus will be on a select few, which demonstrates the
various considerations necessary when evaluating the results. Though,
some general propensities can be seen by observing the results as a
whole.

It is immediately obvious that extrapolations based on
$\Delta_{2\alpha}$ feature a lot less frequently than any of the other
mass relations. The reason is that $\Delta_{2\alpha}$ extrapolates
along the stability curve and towards the super-heavy nuclei. This makes it difficult to compare $\Delta_{2\alpha}$ with any of the other relations as they rarely coincide. From table~\ref{Tab. Ex} it is also evident
that most results are calculated based on only two different mass
relations. Of course, a combination based on additional extrapolations would be preferable, but even if just two are comparable the result's credibility would increase greatly.

The applicability of the mass relations in various areas across the chart of
nuclides is demonstrated by the scattered results. Viewing this
erratic distribution collectively, it is clear that the results are
more consistent with a greater degree of certainty among the heavier
isotopes. The difference between extrapolations with lighter isotopes
are typically $\sim 800 \, \si{\kilo\electronvolt}$, whereas the
heavier isotopes often differ with less than $400 \,
\si{\kilo\electronvolt}$. There are also extrapolations with very large differences, but these are almost exclusively found among the light nuclei. This tendency was to be expected as the
binding energy per nucleon generally varies more for lighter isotopes.

\newpage

\begin{longtable*}{p{0.1\textwidth} *{2}{D{.}{.}{8.3} @{$\pm$} D{.}{.}{5.3}} *{1}{D{.}{.}{7.1} @{$\pm$} D{.}{.}{3.1}} r r} 
\colrule
Nucleus & \multicolumn{4}{c}{Estimates of binding energies [keV]} & \multicolumn{2}{c}{Measured} & \multicolumn{2}{c}{Differences [keV]} \\ 
 \cline{2-5} \cline{8-9}
(Z, N)    &   \multicolumn{2}{c}{Average}  & \multicolumn{2}{c}{Audi and Meng} &  \multicolumn{2}{c}{[keV]} & A $\&$ M  & Exp.\\

\colrule
\endhead

\colrule
\multicolumn{9}{c}{Continued on next page} \\ 
\colrule
\endfoot

\colrule \\
\caption{The average of the extrapolated values
  from table \ref{Tab. Ex} compared to estimates by Audi and
  Meng \cite{Audi} and to experimental measurements of ten nuclei. The differences between the average and both the estimates and the measurements are listed in
  the final two columns. \label{Tab Com}} \endlastfoot

( 26, 43) &577422 &446 &574977 &483 &\multicolumn{2}{c}{-} &$ 2445$ &-\\
( 26, 44) &583561 &583 &580930 &630 &\multicolumn{2}{c}{-} &$ 2631$ &-\\
( 27, 45) &601378 &152 &599688 &576 &\multicolumn{2}{c}{-} &$ 1690$ &-\\
( 28, 46) &624218 &203 &624042 &370 &\multicolumn{2}{c}{-} &$  176$ &-\\
( 32, 55) &719272 &237 &721404 &522 &\multicolumn{2}{c}{-} &$-2132$ &-\\
( 33, 55) &733193 &256 &735328 &440 &\multicolumn{2}{c}{-} &$-2135$ &-\\
( 36, 33) &564201 &1548 &561177 &414 &\multicolumn{2}{c}{-} &$ 3024$ &-\\
( 36, 34) &580395 &192 &578410 &350 &\multicolumn{2}{c}{-} &$ 1985$ &-\\
( 36, 62) &807490 &552 &807324 &490 &\multicolumn{2}{c}{-} &$  166$ &-\\
( 36, 63) &811431 &1659 &809622 &495 &\multicolumn{2}{c}{-} &$ 1809$ &-\\
( 37, 35) &593264 &362 &590328 &504 &\multicolumn{2}{c}{-} &$ 2936$ &-\\
( 38, 64) &846184 &521 &845988 &204 &845904 &70 &$  196$ &$  280$\\
( 40, 38) &642136 &151 &639600 &468 &\multicolumn{2}{c}{-} &$ 2536$ &-\\
( 40, 42) &694614 &145 &694458 &164 &\multicolumn{2}{c}{-} &$  156$ &-\\
( 42, 70) &929068 &271 &928704 &336 &\multicolumn{2}{c}{-} &$  364$ &-\\
( 45, 74) &987949 &193 &988176 &238 &988104 &9 &$ -227$ &$ -155$\\
( 45, 75) &991507 &176 &992160 &240 &\multicolumn{2}{c}{-} &$ -653$ &-\\
( 48, 48) &795033 &1499 &792864 &384 &\multicolumn{2}{c}{-} &$ 2169$ &-\\
( 48, 83) &1083690 &2209 &1075117 &131 &\multicolumn{2}{c}{-} &$ 8573$ &-\\
( 49, 84) &1101543 &4276 &1092861 &266 &\multicolumn{2}{c}{-} &$ 8682$ &-\\
( 50, 51) &835299 &646 &835977 &303 &836391 &300 &$ -678$ &$-1092$\\
( 52, 87) &1140362 &265 &1141607 &417 &1141436 &4 &$-1245$ &$-1074$\\
( 53, 85) &1144358 &355 &1144296 &138 &1144357 &6 &$   62$ &$    1$\\
( 58, 94) &1241501 &738 &1240776 &152 &\multicolumn{2}{c}{-} &$  725$ &-\\
( 60, 68) &1046674 &487 &1046272 &256 &\multicolumn{2}{c}{-} &$  402$ &-\\
( 60, 95) &1266566 &277 &1266505 &155 &1266398 &16 &$   61$ &$  168$\\
( 61, 70) &1069488 &458 &1069222 &131 &\multicolumn{2}{c}{-} &$  266$ &-\\
( 61, 71) &1079277 &303 &1079364 &132 &\multicolumn{2}{c}{-} &$  -87$ &-\\
( 62, 70) &1072254 &122 &1071576 &264 &\multicolumn{2}{c}{-} &$  678$ &-\\
( 62, 71) &1082257 &181 &1081822 &133 &\multicolumn{2}{c}{-} &$  435$ &-\\
( 62, 72) &1094471 &266 &1094244 &134 &\multicolumn{2}{c}{-} &$  227$ &-\\
( 62, 98) &1302401 &214 &1303360 &160 &1303142 &10 &$ -959$ &$ -741$\\
( 63, 72) &1094614 &638 &1094445 &270 &\multicolumn{2}{c}{-} &$  169$ &-\\
( 63, 74) &1116645 &509 &1116550 &137 &\multicolumn{2}{c}{-} &$   95$ &-\\
( 64, 74) &1119820 &109 &1119456 &138 &\multicolumn{2}{c}{-} &$  364$ &-\\
( 64,100) &1332950 &248 &1333484 &328 &\multicolumn{2}{c}{-} &$ -534$ &-\\
( 64,101) &1337784 &295 &1338315 &495 &\multicolumn{2}{c}{-} &$ -531$ &-\\
( 65,100) &1341928 &613 &1341615 &165 &\multicolumn{2}{c}{-} &$  313$ &-\\
( 67,105) &1387762 &639 &1387352 &172 &\multicolumn{2}{c}{-} &$  410$ &-\\
( 68,105) &1397349 &663 &1396802 &173 &\multicolumn{2}{c}{-} &$  547$ &-\\
( 68,106) &1404544 &841 &1403136 &348 &\multicolumn{2}{c}{-} &$ 1408$ &-\\
( 69,108) &1422716 &463 &1422195 &354 &\multicolumn{2}{c}{-} &$  521$ &-\\
( 69,109) &1426908 &361 &1426848 &356 &\multicolumn{2}{c}{-} &$   60$ &-\\
( 70,109) &1436494 &103 &1436475 &358 &\multicolumn{2}{c}{-} &$   19$ &-\\
( 71, 83) &1227341 &438 &1227072 &154 &\multicolumn{2}{c}{-} &$  269$ &-\\
( 71,110) &1450097 &93 &1450172 &362 &1450159 &159 &$  -75$ &$  -62$\\
( 72, 85) &1249835 &187 &1249563 &157 &\multicolumn{2}{c}{-} &$  272$ &-\\
( 73, 85) &1250013 &235 &1249148 &158 &\multicolumn{2}{c}{-} &$  865$ &-\\
( 74, 87) &1272183 &459 &1272061 &161 &\multicolumn{2}{c}{-} &$  122$ &-\\
( 75, 87) &1271759 &302 &1271214 &162 &\multicolumn{2}{c}{-} &$  545$ &-\\
( 75, 92) &1324346 &207 &1324143 &0 &\multicolumn{2}{c}{-} &$  203$ &-\\
( 75,118) &1529776 &345 &1529332 &193 &1529324 &38 &$  444$ &$  452$\\
( 76, 89) &1294078 &229 &1293930 &165 &\multicolumn{2}{c}{-} &$  148$ &-\\
( 77, 88) &1282973 &307 &1283205 &165 &\multicolumn{2}{c}{-} &$ -232$ &-\\
( 77, 93) &1335446 &180 &1335180 &170 &\multicolumn{2}{c}{-} &$  266$ &-\\
( 78, 91) &1315104 &242 &1315327 &169 &\multicolumn{2}{c}{-} &$ -223$ &-\\
( 79, 90) &1303787 &391 &1304004 &338 &\multicolumn{2}{c}{-} &$ -217$ &-\\
( 79, 95) &1356290 &181 &1356852 &174 &\multicolumn{2}{c}{-} &$ -562$ &-\\
( 80, 93) &1336104 &258 &1336252 &173 &\multicolumn{2}{c}{-} &$ -148$ &-\\
( 80,131) &1644115 &447 &1640947 &211 &\multicolumn{2}{c}{-} &$ 3168$ &-\\
( 81, 97) &1377415 &220 &1378076 &178 &\multicolumn{2}{c}{-} &$ -661$ &-\\
( 84,140) &1711704 &352 &1712480 &224 &\multicolumn{2}{c}{-} &$ -776$ &-\\
( 84,141) &1714407 &939 &1716075 &225 &\multicolumn{2}{c}{-} &$-1668$ &-\\
( 84,142) &1720027 &541 &1721216 &452 &\multicolumn{2}{c}{-} &$-1189$ &-\\
( 84,143) &1724127 &237 &1724519 &454 &\multicolumn{2}{c}{-} &$ -392$ &-\\
( 86,144) &1747127 &276 &1747080 &230 &\multicolumn{2}{c}{-} &$   47$ &-\\
( 86,145) &1750834 &393 &1750749 &231 &\multicolumn{2}{c}{-} &$   85$ &-\\
( 88,146) &1772829 &822 &1772784 &468 &1772949 &31 &$   45$ &$ -120$\\
( 90,147) &1792150 &232 &1792194 &474 &\multicolumn{2}{c}{-} &$  -44$ &-\\
( 92,149) &1816465 &247 &1816899 &241 &\multicolumn{2}{c}{-} &$ -434$ &-\\
( 98,140) &1786909 &237 &1787142 &476 &\multicolumn{2}{c}{-} &$ -233$ &-\\
(100,144) &1821979 &172 &1822192 &244 &\multicolumn{2}{c}{-} &$ -213$ &-\\

\end{longtable*}

The uncertainty connected to the final average value mostly comes from
only one of the relevant extrapolations. This indicates that the extrapolations generally are
compatible, and makes the final results more credible. However,
some results are questionable, where the uncertainty has been
calculated based on the distance between extrapolations.  For instance
is $\sigma > 1 \, \si{\mega\electronvolt}$ for $(48, 48)$, $(48, 83)$, $(49, 84)$,
and $(52, 52)$ which makes these results less useful. Such
uncertainties are not surprising in view of the involved isotopes,
where for example rapidly varying shell effects are pronounced. On the other hand, the significant uncertainty of for instance $(84, 141)$ is somewhat more troubling. Considering the involved extrapolations a better result could have been expected. This goes to show the volatility the method in the vicinity of shells, and emphasizes the care that must be taken when analysing these results.

Inconsistent results like those are clearly in the minority, as most
have an acceptable uncertainty based on very compatible
extrapolations. For instance the averages for $(84, 98)$, $(86,
103)$, $(86, 144)$, $(90, 116)$, $(94, 132)$ and $(94, 154)$ are all
based on very consistent extrapolations. The benefit of combining
different mass relations is also emphasized when considering $(90,
116)$, where $\Delta_{2n}$ and $\Delta_{2p}$ have extrapolations with
significant uncertainties. The average is then based on
$\Delta_{2\alpha}$ and $\Delta_{2(N-Z)}$, but the final average value is
actually consistent with the extrapolations based on $\Delta_{2n}$ and
$\Delta_{2p}$.

Results where $Z > 82$ and $N < 126$ are perhaps more interesting. The
area defined by these shells has traditionally been difficult to
estimate, and isotopes in this area extrapolated by multiple mass
relation deserves special attention. Some of these extrapolations
involve shells, and must be viewed with suspicion. The nine
extrapolations in this area are generally internally consistent,
including even those influenced by shells. They must be used with
care, but the remaining majority seem to be especially reliable.  In
particular, the results for $(90, 116)$ and $(90, 117)$ are based on
very close-lying extrapolations, and $(90, 116)$ also have a very
reasonable uncertainty.

Actually, it could be argued that the procedure is too exclusive in
certain situations. For instance the heavier isotopes like $(90,
117)$, or in particular $(98, 140)$, and $(100, 144)$, could possibly
have included additional mass relations in the calculations. Here some
extrapolations have been excluded based on their uncertainties, even
though the final result agrees almost perfectly with these
extrapolations. The preferred attitude has been to err on the side of
caution, which is why these extrapolations have been excluded.

To get an indication of whether the averages are reasonable
extrapolations, they are in table \ref{Tab Com} compared with
estimates provided by Audi and Meng \cite{Audi}. For $Z
< 64$ the deviation between the estimates are often greater than $1 \,
\si{\mega\electronvolt}$, which again indicates that extrapolations
are less reliable for light isotopes.  However, for $Z > 64$ the
differences are usually less than $300 \, \si{\kilo\electronvolt}$, and
the uncertainties are also very much comparable. 

Table \ref{Tab Com} also includes a comparison with ten nuclei not included in Audi and Meng's preliminary mass table. The measurement of $^{139}$Te was done by Hakala
\textit{et al} \cite{Hakala}, whereas $^{138}$I, $^{155}$Nd and
$^{160}$Sm was measured by Van Schelt \textit{et al}
\cite{Schelt}. The last six isotopes $^{101}$Sn, $^{102}$Sr, $^{119}$Rh, $^{181}$Lu, $^{193}$Re, and $^{234}$Ra are included in the final version of Ame2012 \cite{Ame2012}. There is a significant difference between the
extrapolations of $^{101}$Sn, $^{139}$Te and $^{160}$Sm and the measured values. The
extrapolations in these cases includes either the shell $Z = 50$ or
the subshell $Z = 64$, which could explain the large deviation. Shell effects could also explain the not insignificant deviation of $^{193}$Re. On the
other hand, the extrapolations of $^{102}$Sr, $^{119}$Rh, $^{138}$I, $^{155}$Nd, $^{181}$Lu, and $^{234}$Ra are very
much comparable to the measured values. Despite the fact that 138-I
are very close to both the $Z = 50$ and the $N = 82$ shell, the
extrapolation predicted exactly the value measured. Apparently, the
method can at times be applied near magic numbers.

\begin{figure}
\centering
	\includegraphics[width=\columnwidth]{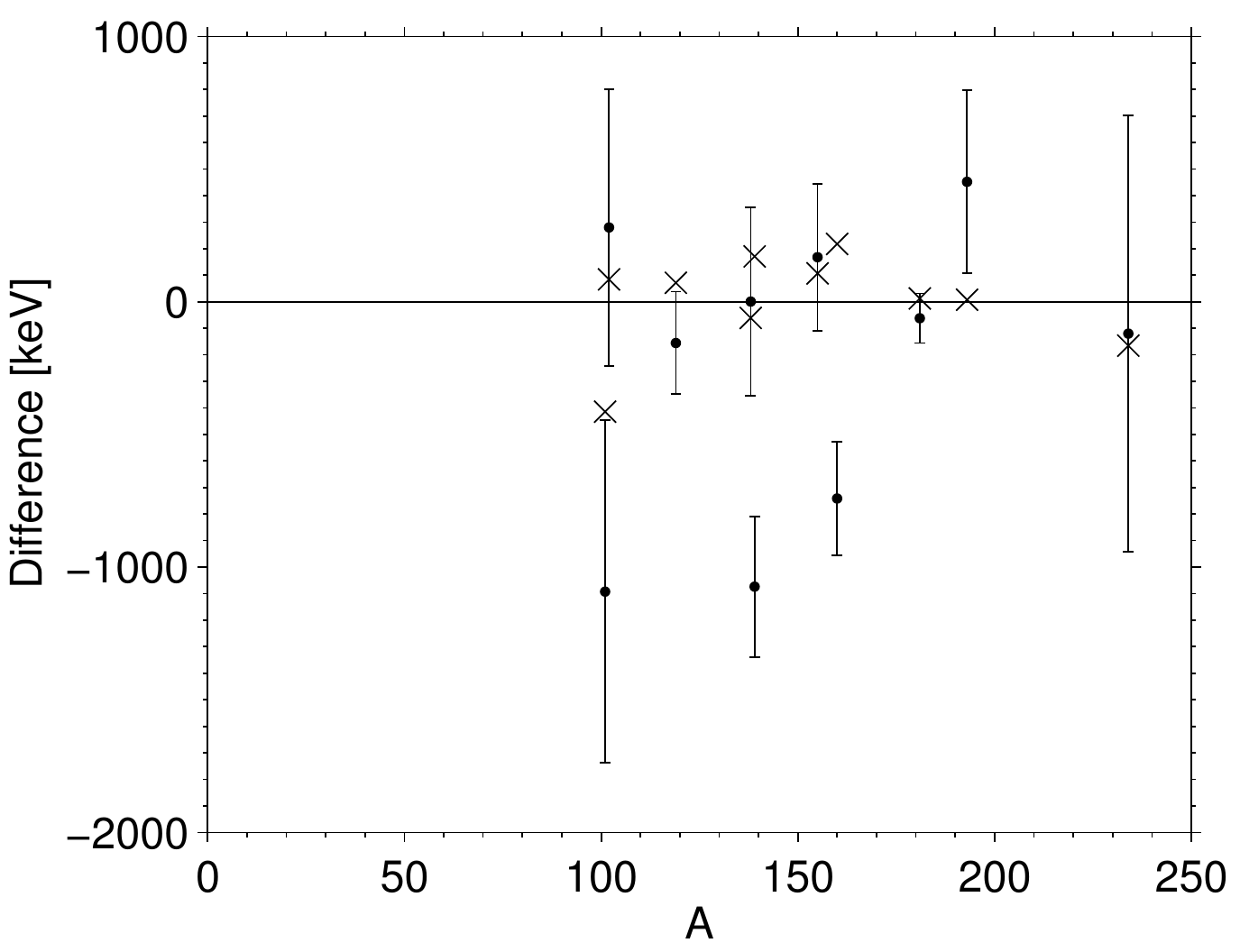}
\caption{The differences between our extrapolations and the measured values where the errorbars are based solely on the extrapolated uncertainties. The crosses indicate the difference between Audi and Meng's extrapolations and the measurements.  \label{Fig. Diff}}
\end{figure}

Fig. \ref{Fig. Diff} provides an overview of the differences between the extrapolated and the measured values as a function of nucleon number. The three nuclei deviating by more than $500$~keV in our extrapolation are influenced by closed shell or subshell effects which add to the inaccuracy in the present type of extrapolation. The other more believable points deviate on average by about $200$~keV. 

The corresponding extrapolations by Audi and Meng exhibit a comparable deviation, on average about $130$~keV. As seen from table \ref{Tab Com} their uncertainties are also comparable. However, shell effects seem included in their extrapolations.

\section{Conclusion \label{Conclusion}} 

The ultimate purpose of this paper was to extrapolate new binding
energies, using several mass relations constructed specifically to
this task.

A very general model for describing the binding energy was assumed
based on known contributions. Four relations were then designed to
eliminate as many factors as possible in the description of the
binding energy. The intent was to identify groupings of isotopes where
the mass relation either cancelled completely or showed clear,
predictable tendencies. By continuing these tendencies outcomes could
be predicted, and the binding energy of unknown isotopes could be
extrapolated accordingly.

Four mass relations were defined and applied individually. The
results were used to confirm the predictions of cancellation of
smoothly vanishing aspects, and by extension to corroborate the initial
assumption of dividing the binding energy in qualitatively different
terms. Each of the mass relations supplied numerous extrapolations,
which were scattered across the chart of nuclides. This scattering
demonstrated that the applicability of the method was not limited to a
specific area, though the results were generally more reliable with
heavier isotopes. Many isotopes were also extrapolated by several mass
relations, which provided several comparable estimates for the given
isotopes. In addition, it allowed for a combined result based on
extrapolations from different mass relations.

When comparing or combining different extrapolations some
considerations had to be made. Some extrapolations had too large
uncertainties and were excluded from any calculations. These
extrapolations were considered too unreliable and would not improve
the final result. If significant discrepancies were found when
calculating the average, any extrapolation influenced by effects known
to be significant was excluded as well. 

Several extrapolations could be combined with these considerations in
mind, and the final results were displayed in tables where several
expected general tendencies were observed. In particular, results
regarding heavier isotopes were more consistent and more reliable.
The unique orientation of one mass relation along the stability
curve resulted in fewer possible extrapolations comparable with the
other three relations. 

Comparisons of extrapolated results, based on any of the four mass
relations, in general showed rather close agreement. Even when an
extrapolation was discarded based on its uncertainty it was often in
close vicinity to the final average. The extrapolations at higher
nucleon number usually differed by at most $\sim 400 \,
\si{\kilo\electronvolt}$. These averages were consistent with other
estimates, and the uncertainties were of the same magnitude. On the
other hand, the averages seemed much less accurate for lighter
isotopes, and the method is probably not competitive for these
isotopes.

A number of the results in the region where $Z > 82$ and $N < 126$
were acceptable both in consistency and uncertainty. This is
particularly interesting given that this region traditionally is very
difficult to estimate. Some results, notably in this area, were
influenced by shell effects, and should be handled with care, but
generally the calculated averages seemed to be reliable.

Ten isotopes estimated by combining extrapolations were measured
after Audi and Meng complied their initial information. This allowed for a
direct evaluation of the accuracy of the method. The values for most extrapolations corresponded very well to the extrapolations.
Those that deviated significantly were all in the vicinity of shells or
subshells, and the inaccuracy of the method in such areas is not
surprising, as no attempt has been made to account for these
effects. On the contrary, it is more surprising that the extrapolation
of $^{138}$I is so accurate as this isotope is also close to shells.

The greatest fundamental weakness with the presented method is the use of somewhat removed isotopes. Isotopes are
combined over a significant distance, particularly when calculating
the variation in the tendencies. Combining isotopes over a greater
distance increases the likelihood of combining unrelated effects. It
is difficult to account for the effect of one isotope being influenced
differently than the others in the extrapolation. The usual approach
has been to apply more compact mass relations. Beginning with the
Garvey-Kelson mass relations, this fear of combining unrelated effects
has been an ongoing concern. However, despite the rather large span
of the mass relations used here, the results are comparable to the
best of other available extrapolations.

Although, applying other, possibly more compact, mass relations would
be the most obvious way to supplement the results. By combining
results from multiple mass relations, and not just the four applied
here, this method also allows for convenient extensions and
improvements. Applying more complex mass relations, could possibly
increase the applicability of this method even more. By creating a
comprehensive system of extrapolations based on different relations it
would probably be possible to determine binding energies with still
greater precision, and in greater number. The actual binding energy
would be approached from many directions by several mass relations, and
the final result would be all the more credible.

Even though the extrapolated binding energies might not be perfectly
consistent, the calculated averages should still be very viable and
useful estimates. In particular, results for isotopes unencumbered by
shell effects and the like should be more than reliable.

Finally, from the evaluations of $Q$-values along the stability line
it was possible to examine general structures in the binding energy of
super-heavy isotopes. Here signature of a minor neutron shell at $N =
152$ was found. Applying the mass relation perpendicular to the
stability curve the behaviour across this neutron number was found to
be characteristic of a closed shell. These findings very clearly
suggests there exists a minor neutron shell at $N = 152$.  It is also
striking that no other shell is revealed in this region which owes its
very existence to stability provided by shell effects.

In conclusion, simple four-nucleus mass relations, where smooth
contributions to the nuclear binding energy vanish to second order,
are used to extrapolate unknown nuclear binding energies with rather good
accuracy. We provide estimates for a series of different nuclei just
outside the region of knowledge where a good deal of present nuclear
research activities are focussed. In particular, we apply the method
to the super-heavy region where special $Q$-values are measured very
accurately.


\end{document}